\documentclass[aps,prd,amsmath,amssymb,twocolumn,preprintnumbers]{revtex4-2}
\usepackage[english]{babel}
\usepackage{bm}
\usepackage{bbm}
\usepackage{mathtools}
\usepackage{graphicx}
\usepackage{tikz}
\usetikzlibrary{arrows,shapes,backgrounds,calc,positioning,tikzmark,patterns,automata,decorations.markings}
\tikzset{fontscale/.style={font=\relsize{#1}}}
\tikzset{->-/.style={decoration={
  markings,
  mark=at position #1 with {\arrow{>}}},postaction={decorate}}}
\tikzset{-<-/.style={decoration={
  markings,
  mark=at position #1 with {\arrow{<}}},postaction={decorate}}}

\tikzset{cross/.style={cross out,draw,minimum size=2*(#1-\pgflinewidth),inner sep=0pt, outer sep=0pt}}

\tikzset{
  pics/carc/.style args={#1:#2:#3}{
    code={
      \draw[pic actions] (0,0) -- (#1:#3) arc(#1:#2:#3) -- cycle;
    }
  }
}
\usepackage{picture}
\usepackage{cancel}
\usepackage{tabularx}
\usepackage{tensor}
\usepackage{float}
\usepackage{fancyhdr}
\usepackage{xcolor}
\usepackage{scalerel}
\usepackage{relsize}
\usepackage[absolute,overlay]{textpos}
\usepackage[customcolors]{hf-tikz}
\usepackage{cals}
\usepackage{enumitem}
\setlist{nolistsep}
\usepackage{pifont}
\usepackage[export]{adjustbox} % adds left,center,right as options for \includegraphics
\usepackage[most]{tcolorbox}
\usepackage{varwidth}
\usepackage{ragged2e}
\usepackage{ifthen} % provides \ifthenelse test  
\usepackage{xifthen}
\usepackage{caption}
\DeclareCaptionLabelFormat{cont}{#1~#2\alph{ContinuedFloat}}
\DeclareCaptionLabelFormat{custom}
{%
      \textbf{#1~#2:}
}
\DeclareCaptionLabelSeparator{custom}{}
\DeclareCaptionFormat{custom}
{%
    #1#2 \justifying #3
}
\usepackage[longend,linesnumbered]{algorithm2e}
\let\oldnl\nl% Store \nl in \oldnl
\newcommand{\nonl}{\renewcommand{\nl}{\let\nl\oldnl}}% Remove line number for one line
\SetAlCapSkip{5pt}
\SetAlCapHSkip{3pt}
\makeatletter
\renewcommand{\algocf@captiontext}[2]{%
\noindent#1\textbf{:}
\AlCapFnt{}#2
}%
\renewcommand{\algocf@makecaption@plain}[2]{%
  \global\sbox{\algocf@capbox}{\hskip\AlCapHSkip%
    \setlength{\hsize}{\columnwidth}% restored on exit of sbox
    \addtolength{\hsize}{-2\AlCapHSkip}% add equal margin to both sides
    \parbox[t]{\hsize}{\justifying\algocf@captiontext{#1}{#2}}}% then caption is not centered
}%
\makeatother

\tcbuselibrary{skins}
\usepackage[pdfborder={0 0 0}]{hyperref}
\definecolor{darkblue}{rgb}{0,0,.5}
\definecolor{darkgreen}{rgb}{0,0.5,.5}
\definecolor{darkyellow}{rgb}{0.5,0.5,0}
\definecolor{fhl}{rgb}{1,0,0}
\hypersetup{colorlinks=true, breaklinks=true, linkcolor=darkblue, menucolor=darkblue, urlcolor=darkblue, linktocpage=true}

\newtcolorbox{hlbox}[2][red]{
colbacktitle=#1!10,
colback=white!95!#1,
coltitle=black,
fonttitle=\bfseries,
colframe=#1!50,
boxrule=0.5pt,
titlerule=0pt,
title={\strut#2},
arc=3pt,
middle=0pt,
boxsep=0pt,
left skip=0pt,
right skip=0pt}

\makeatletter

\newsavebox\myboxA
\newsavebox\myboxB
\newlength\mylenA

\newcommand*\xoverline[2][0.75]{%
    \sbox{\myboxA}{$\m@th#2$}%
    \setbox\myboxB\null% Phantom box
    \ht\myboxB=\ht\myboxA%
    \dp\myboxB=\dp\myboxA%
    \wd\myboxB=#1\wd\myboxA% Scale phantom
    \sbox\myboxB{$\m@th\overline{\copy\myboxB}$}%  Overlined phantom
    \setlength\mylenA{\the\wd\myboxA}%   calc width diff
    \addtolength\mylenA{-\the\wd\myboxB}%
    \ifdim\wd\myboxB<\wd\myboxA%
       \rlap{\hskip 0.5\mylenA\usebox\myboxB}{\usebox\myboxA}%
    \else
        \hskip -0.5\mylenA\rlap{\usebox\myboxA}{\hskip 0.5\mylenA\usebox\myboxB}%
    \fi}

\makeatother

\let\originalleft\left
\let\originalright\right
\renewcommand{\left}{\mathopen{}\mathclose\bgroup\originalleft}
\renewcommand{\right}{\aftergroup\egroup\originalright}
\newcommand{\e}{\operatorname{e}}
\newcommand{\SU}[1]{\operatorname{SU}\left(#1\right)}

\newcommand{\su}[1]{\mathfrak{su}\left(#1\right)}

\newcommand{\of}[1]{\left(#1\right)}

\newcommand{\bof}[1]{\biggl(\bigg.#1\bigg.\biggr)}
\newcommand{\sof}[1]{\bigl(\big.#1\big.\bigr)}
\newcommand{\ssof}[1]{(#1)}

\newcommand{\bfof}[1]{\biggl[\bigg.#1\bigg.\biggr]}
\newcommand{\sfof}[1]{\bigl[\big.#1\big.\bigr]}
\newcommand{\ssfof}[1]{[#1]}
\newcommand{\cof}[1]{\left\{\right.#1\left.\right\}}

\newcommand{\floor}[1]{\lfloor #1 \rfloor}
\newcommand{\ceil}[1]{\lceil #1 \rceil}

\newcommand{\trace}{\operatorname{tr}}
\renewcommand*{\det}{\mathop{\mathrm{det}}\nolimits}

\newcommand{\avof}[1]{\left\langle #1\right\rangle}

\newcommand{\repart}{\operatorname{Re}}

\newcommand{\ii}{\mathrm{i}}

\newcommand{\dd}{\mathrm{d}}

\newcommand{\totd}[3]{\frac{\dd^{#3} #1}{\dd #2^{#3}}}

\newcommand{\partd}[2]{\frac{\partial #1}{\partial #2}}

\newcommand{\order}[1]{\mathcal{O}\big(#1\big)}

\newcommand{\id}{\mathbbm{1}}

\newcommand{\abs}[1]{\left| #1\right|}
\newcommand{\ssabs}[1]{| #1|}

\renewcommand*\[{\begin{equation}}
\renewcommand*\]{\end{equation}}

\renewcommand*\bar[1]{\ThisStyle{\xoverline{\SavedStyle #1}}}

\renewcommand*\hat[1]{\widehat{#1}}
\let\oldstackrel\stackrel
\renewcommand*\stackrel[2]{{\scriptstyle\oldstackrel{#1}{#2}}}

\definecolor{emphcol}{rgb}{1.,0,0}
\newcommand{\cemph}[2][]{%
    \emph{\ifthenelse{\isempty{#1}}{\textcolor{emphcol}{#2}}{\textcolor{#1}{#2}}}
}

\newcommand{\psmatrix}[1]{\left(\begin{smallmatrix}#1\end{smallmatrix}\right)}

\makeatletter
\pgfarrowsdeclare{open cap}{open cap}
{\pgfarrowsleftextend{+0pt}\pgfarrowsrightextend{+0.5\pgflinewidth}}
{
  \pgfmathsetlength{\pgfutil@tempdimb}{.5*\pgflinewidth-.5*\pgfinnerlinewidth}%
  \pgfsetlinewidth{\pgfutil@tempdimb}
  \pgfsetbuttcap
  \pgfsetdash{}{0pt}
  \pgfmathsetlength{\pgfutil@tempdima}{.5*\pgfutil@tempdimb+.5*\pgfinnerlinewidth}%
  \pgfpathmoveto{\pgfqpoint{0pt}{\pgfutil@tempdima}}
  \pgfpatharc{270}{90}{-\pgfutil@tempdima}
  \pgfusepathqstroke
}
\makeatother
\allowdisplaybreaks

\begin{document}
\title{Evaluating matrix power series with the Cayley-Hamilton theorem}

\author{Tobias Rindlisbacher}
\email{tobias.rindlisbacher@helsinki.fi}
\affiliation{Department of Physics \& Helsinki Institute of Physics, University of Helsinki, P.O. Box 64, FI-00014 University of Helsinki, Finland}
%\affiliation{Albert Einstein Center for Fundamental Physics \& Institute for Theoretical Physics, University of Bern Sidlerstrasse 5, CH-3012 Bern, Switzerland}

\begin{abstract}
The Cayley-Hamilton theorem is used to implement an iterative process for the efficient numerical computation of matrix power series and their differentials. In addition to straight-forward applications in lattice gauge theory simulations e.g. to reduce the computational cost of smearing, the method can also be used to simplify the evaluation of $\SU{N}$ one-link integrals or the computation of $\SU{N}$ matrix logarithms. 
\end{abstract}
\maketitle

\section{Introduction}\label{sec:intro}
In lattice field theory, the Cayley-Hamilton theorem is most well known from studies of $\SU{N}$ gauge-fermion theories that make use of some sort of gauge smearing to reduce UV cutoff effects. In the stout~\cite{Morningstar:2003gk} and HEX~\cite{Capitani:2006ni} smearing schemes, for example, the theorem can be used to perform the required matrix exponential computations, and in the nHYP smearing scheme~\cite{Hasenfratz:2001hp} to evaluate (inverse) matrix square roots~\cite{Hasenfratz:2007rf,DeGrand:2012qa,DeGrand:2016pur}. Beyond applications in smearing, the theorem has been used to compute matrix logarithms in gauge fixing procedures~\cite{Hudspith:2014rma}, and more recnetly, to perform simulations with an improved lattice Dirac operator that involves matrix exponentials~\cite{Francis:2019muy}. 

In most of these studies the Cayley-Hamilton theorem has, however, been applied "manually", in the sense that it has been used to derive algorithms which efficiently perform a given operation, e.g. matrix exponentiation, for a given matrix size. If the matrix size changes, e.g. because one intends to study a theory with $\SU{5}$ instead of $\SU{3}$ gauge group, the algorithm needs to be adjusted. An exception to this is the application of the Cayley-Hamilton theorem as described in~\cite{DeGrand:2016pur}. There it is used that the theorem can be applied by solving numerically a matrix equation that involves the Vandermonde matrix of eigenvalues of the input matrix. This latter approach is in principle applicable for any matrix size, but requires some care to deal with cases where the Vandermonde matrix becomes singular or nearly singular.  

In the present work we are going to discuss a different approach which will not require the computation of matrix eigenvalues. Instead, the Cayley-Hamilton theorem is used to derive an iterative process for the efficient numerical evaluation of matrix power series of arbitrary square matrices $U$. If needed, the iterative process allows also for a simultaneous computation of the differential of the matrix power series of $U$, respectively, of its derivatives with respect to the components of the matrix $U$.

The basic idea behind our method has briefly been discussed already in a review of computational methods for coumputing matrix expnentials about 20 years ago~\cite{Cleve:2006mfk}. There it was pointed out that the algorithm is not particularly favorable because of its computational complexity being $\order{N^4}$ in matrix size $N$, and it potential instability. In this work we discuss a regularized version of the algorithm and present results showing that up to $N=10$ it outperforms other algorithms for computing matrix exponentials with respect to speed and accuracy.     

The paper is organized as follows. Sec.~\ref{sec:method} provides a derivation  of the iterative Cayley-Hamilton method for matrix power series and discusses some possible extensions. Sec.~\ref{sec:implementation} describes a possible implementation of the method and how it can be extended to simultaneously compute also the differentials of a matrix power series. In Sec.~\ref{sec:applications} the method is applied to the problem of computing matrix exponentials and to simplify the numerical computation of $\SU{N}$ matrix logarithms and $\SU{N}$ one link integrals. In Sec.~\ref{sec:numtests} we compare performance and numerical accuracy of our method for computing matrix exponentials with those of other methods and present some results. Sec.~\ref{sec:summary} provides a summary.

\section{Iterative Cayley-Hamilton}\label{sec:method}
Consider a matrix $U\in\mathbb{C}^{N\times N}$ on which  we would like to evaluate the matrix power series that is obtained by replacing in the Taylor series of some scalar function $f\of{x}$,
\[
f\of{x}=\sum\limits_{n=0}^{\infty}\,r_{n}\,x^{n}\ ,\label{eq:scalfunc}
\]
the powers of the scalar variable $x$ by the corresponding matrix powers of $U$. In the following, we will use the notation $f\of{U}$ to refer to this procedure. 

A naive evaluation of $f\of{U}$ is computationally expensive, since it involves a large number of matrix multiplications and summations. A more efficient way of evaluating $f\of{U}$ would be to perform an eigen-decomposition of $U$ and apply $f\of{x}$ to the eigenvalues in the diagonal factor of the decomposition. However, if also the derivatives of $f\of{U}$ with respect to the components of $U$ are required, this approach can become tedious since the factors of numerically performed eigen-decompositions are in general not smooth functions of the components of $U$. The latter problem can be overcome by the already mentioned approach descried in~\cite{DeGrand:2016pur}, at the cost of having to solve Vandermonde matrix equations.

In the following we describe an alternative approach in the form of an iterative method derived from the Cayley-Hamilton theorem, which allows for an efficient computation of $f\of{U}$ and also of its derivatives with respect to the components of $U$.

\subsection{Basic method}\label{ssec:basicmethod}
The Cayley-Hamilton theorem states that if $U$ is an $N\times N$ matrix, then any power $U^n$ with $n\geq N$ can be written as a finite superposition of powers $U^i$ with $i\in\cof{0,\ldots,N-1}$, i.e.
\[
U^n=\sum_{i=0}^{N-1}\,a_{n,i}\,U^i\ .\label{eq:monodecomp}
\]

In particular, if $n=N$, then the coefficients $a_{N,i}$, $i\in\cof{0,\ldots,N-1}$ are given by $a_{N,i}=-c_{i}$ $\forall\,i\in\cof{0,\ldots,N-1}$ with $\cof{c_{i}}_{i\in\cof{0,\ldots,N}}$ being the set of coefficients of the characteristic polynomial $p$ of $U$:
\[
p\of{\lambda}=\det\of{\id\,\lambda-U}=\sum_{i=0}^{N}\,c_i\,\lambda^{i}\ .\label{eq:chpoly}
\]

We now observe, that if for some $n$ the coefficients $\cof{a_{n,i}}_{i\in\cof{0,\ldots,N-1}}$ are known, then we can compute $\cof{a_{n+1,i}}_{i\in\cof{0,\ldots,N-1}}$ by noting that
\[
U^{n+1}=U^{n}\,U
\]
which after applying \eqref{eq:monodecomp} to $U^{n+1}$ and $U^{n}$ reads:
\[
\sum_{i=0}^{N-1}\,a_{n+1,i}\,U^{i}=\sum_{i=0}^{N-1}\,a_{n,i}\,U^{i+1}\ .\label{eq:recurrcond}
\]
In equation \eqref{eq:recurrcond} the last term in the sum on the right-hand side of the equality sign (for which $i+1=N$) can be replaced by  
\[
U^N=-\sum_{i=0}^{N-1}\,c_{i}\,U^i\ ,\label{eq:monodecompcharpoly}
\]
after which a comparison of coefficients between terms of equal powers in $U$ on the left- and right-hand side of the equality sign in \eqref{eq:recurrcond} leads to the following recurrence relations in $n$:
\begin{subequations}\label{eq:ankparam}
\begin{align}
a_{n,0}&=-a_{n-1,N-1}\,c_{0}\ ,\label{eq:ankparama}\\
a_{n,k}&=a_{n-1,k-1}-a_{n-1,N-1}\,c_{k}\ ,\label{eq:ankparamb}
\end{align}
for $n\geq N$, $0< k\leq N-1$, and the trivial low order coefficients for $0\leq n<N$, $0\leq k\leq N-1$, given by
\[
a_{n,k}=\delta_{n,k}\ ,\label{eq:ankparamc}
\]
\end{subequations}
with $\delta_{n,k}$ being the Kronecker delta, serve as initial conditions. Note that each iteration step requires only $\order{N}$ operations.

By viewing the Cayley-Hamilton coefficients $a_{n,i}$, $i\in\cof{0,\ldots,N-1}$ at each order $n\in\mathbb{N}_0$ as $N$-component column vectors,
\[
v_n=\of{a_{n,i}}_{i\in\cof{0,\ldots,N-1}}\quad\forall n\in\mathbb{N}_0\ ,
\]
the iteration rule \eqref{eq:ankparam} can be written as
\[
v_{n}=A\,v_{n-1}\qquad \forall n\in\mathbb{N}\ ,\label{eq:vectorchiter}
\]
where the $N\times N$ matrix $A$ is the so-called companion matrix for the characteristic polynomial, Eq.~\eqref{eq:chpoly}, given by
\[
A=\psmatrix{
  &   &        &   &   & -c_0\\
1 &   &        &   &   & -c_1\\
  & 1 &        &   &   & -c_2\\
\phantom{\ddots}  & \phantom{\ddots}  & \ddots & \phantom{\ddots}  & \phantom{\ddots}  & \vdots\\[6pt]
  &   &        & 1 &   & -c_{N-2}\\
  &   &        &   & 1 & -c_{N-1}\\}\ .\label{eq:companionmat}
\]

Equipped with the set of Cayley-Hamilton coefficients $\cof{a_{n,i}}_{n\in\mathbb{N}_0,i\in\cof{0,\ldots,N-1}}$ for the matrix $U$, the function $f\of{x}$ from \eqref{eq:scalfunc} applied to $U$ can be written as:
\begin{multline}
f\of{U}=\sum\limits_{n=0}^{\infty}\,r_{n}\,U^{n}=\sum\limits_{n=0}^{\infty}\,r_{n}\,\sum\limits_{i=0}^{N-1}\,a_{n,i}\,U^{i}\\
=\sum\limits_{i=0}^{N-1}\bof{\underbrace{\sum\limits_{n=0}^{\infty}\,r_{n}\,a_{n,i}}_{\bar{r}_{i}}}\,U^{i}=\sum\limits_{i=0}^{N-1}\,\bar{r}_{i}\,U^{i}\ .\label{eq:chfresult}
\end{multline}

\subsection{Extensions}\label{ssec:extensions}
After having discussed the basic working principle of the iterative Cayley-Hamilton method for evaluating matrix power series, it is worth point out some possible extensions to the basic use case.

\subsubsection{Simultaneous evaluation of multiple scalar functions}\label{sssec:simuevalmultfunc}
Since the coefficients $\cof{a_{n,i}}_{n\in\mathbb{N}_0,i\in\cof{0,\ldots,N-1}}$ in \eqref{eq:chfresult} are mere functions of the characteristic polynomial coefficients $\cof{c_i}_{i\in\cof{0,\ldots,N-1}}$ of $U$, and do not depend on the function $f\of{x}$, the same $\cof{a_{n,i}}_{n\in\mathbb{N}_0,i\in\cof{0,\ldots,N-1}}$ can be used to evaluate multiple functions $f_{\alpha}\of{x}=\sum_{n=0}^{\infty}r_{\alpha,n}\,x^n$, $\alpha=1,2,\ldots$ simultaneously on $U$; one simply needs to plug the corresponding polynomial coefficients $\cof{r_{\alpha,n}}_{n\in\mathbb{N}_0}$ in \eqref{eq:chfresult}. 

This can for example be used to efficiently evaluate $f\of{s\,U}$ for a sequence of different $s\in\mathbb{C}$, as one has:
\begin{multline}
f\of{s\,U}=\sum\limits_{n=0}^{\infty}\,r_{n}\,\of{s\,U}^{n}=\sum\limits_{n=0}^{\infty}\,\underbrace{r_{n}\,s^{n}}_{r_{n}\of{s}}\,U^{n}\\
=\sum\limits_{i=0}^{N-1}\bof{\underbrace{\sum\limits_{n=0}^{\infty}\,r_{n}\of{s}\,a_{n,i}}_{\bar{r}_{i}\of{s}}}\,U^{i}=\sum\limits_{i=0}^{N-1}\,\bar{r}_{i}\of{s}\,U^{i}\ .\label{eq:chfsresult}
\end{multline}
For given $\cof{a_{n,i}}_{n\in\mathbb{N}_0,i\in\cof{0,\ldots,N-1}}$, the computational cost of evaluating the coefficients $\cof{\bar{r}_{i}\of{s}}_{i\in\cof{0,\ldots,N-1}}$ for a given $s\in\mathbb{C}$ is about $N$ times the cost of evaluating $f\of{s}$, and the cost for composing the final matrix $f\of{s\,U}$ from $\cof{\bar{r}_{i}\of{s}}_{i\in\cof{0,\ldots,N-1}}$ and the pre-computed $\cof{U^i}_{i\in\cof{0,\ldots,N-1}}$ corresponds to $\sim N^3$ multiplication and addition operations. 

\subsubsection{Derivatives and differentials}\label{sssec:derivatives}
A simple derivative formula follows directly from \eqref{eq:chfsresult}, namely that
\[
\totd{}{s}{}f\of{s\,U}=\sum\limits_{i=0}^{N-1}\,\bar{r}'_{i}\of{s}\,U^{i}\ ,
\]
where the prime refers to differentiation with respect to $s$, and one has,
\[
\bar{r}'_{i}\of{s}=\sum\limits_{n=1}^{\infty} r'_{n}\of{s}\,a_{n,i}\quad\forall i\in\cof{0,\ldots,N-1}\ .
\]

The iterative Cayley-Hamilton method allows, however, also for the computation of general differentials
\[
\dd f\of{U}=\partd{f\of{U}}{U\indices{^a_b}}\,\dd  U\indices{^a_b}\ ,
\]
along with the evaluation of $f\of{U}$ itself. This will be discussed in detail in Sec.~\ref{ssec:computingderivterms}.

\subsubsection{Negative powers}\label{sssec:negpowers}
In the case where the matrix $U$ is invertible, the coefficient $c_0$ of the characteristic polynomial of $U$ is non-zero and \eqref{eq:ankparam} can be reversed to compute the coefficients for \eqref{eq:monodecomp} with $n<0$, i.e. to compute negative powers of the matrix $U$. The recurrence relations then read for $n<-1$ and $0\leq k<N-1$:
\begin{subequations}\label{eq:anegnkparam}
\begin{align}
a_{n,N-1}&=-a_{n+1,0}/c_{0}\ ,\\
a_{n,k}&=a_{n+1,k+1}-a_{n+1,0}\,c_{k+1}/c_{0}\ ,
\end{align}
with the initial conditions for $n=-1$ given by
\[
a_{-1,k}=-c_{k+1}/c_{0}\ ,\ 0\leq k <N \ .
\]
\end{subequations}
The latter are obtained by multiplying both sides of \eqref{eq:monodecompcharpoly} by $U^{-1}$ and then solving for $U^{-1}$, or by simply inverting \eqref{eq:vectorchiter}.

\subsubsection{Product rule}\label{sssec:productrule}
Assume we have computed the Cayley-Hamilton decomposition coefficients for a family of matrix functions $f_k\of{U}$, $k=1,2,\ldots$ applied to a $N\times N$ matrix $U$, so that we can write:
\[
f_k\of{U}=\sum\limits_{i=0}^{N-1}\,\bar{r}_{k,i}\,U^{i}\ .
\]
We can then compute the Cayley-Hamilton decomposition of the product of any pair $\of{f_{k}\of{U}, f_{l}\of{U}}$ directly from the sets of coefficients $\cof{\bar{r}_{k,i}}_{i\in\cof{0,\ldots,N-1}}$, $\cof{\bar{r}_{l,i}}_{i\in\cof{0,\ldots,N-1}}$ and the Cayley-Hamilton coefficients $a_{n,i}\of{U}$ for $n\in\cof{0,\ldots,2\,\of{N-1}}$, $i\in\cof{0,\ldots,N-1}$. We start by writing
\begin{multline}
f_{\of{k,l}}\of{U}=f_{k}\of{U}\,f_{l}\of{U}=\sum\limits_{i,j=0}^{N-1}\,\bar{r}_{k,i}\,\bar{r}_{l,j}\,\underbrace{U^{i+j}}_{\mathclap{\sum\limits_{m=0}^{N-1}\,a_{i+j,m}\,U^{m}}}\\
=\sum\limits_{m=0}^{N-1}\bof{\underbrace{\sum\limits_{i,j=0}^{N-1}\bar{r}_{k,i}\,\bar{r}_{l,j}\,a_{i+j,m}}_{\bar{r}_{\of{k,l},m}}}\,U^{m}\ ,
\end{multline}
and note that $a_{i+j,m}$ can for $i,j\in\cof{0,\ldots,N-1}$ be written in terms of the companion matrix \eqref{eq:companionmat} as:
\[
a_{i+j,m}=\ssof{A^i}\indices{^m_j}=\ssof{A^j}\indices{^m_i} .
\]
It follows that the Cayley-Hamilton decomposition coefficients $\cof{\bar{r}_{\of{k,l},m}}_{m\in\cof{0,\ldots,N-1}}$ of the product $f_{\of{k,l}}\of{U}=f_{k}\of{U}\,f_{l}\of{U}$ can be computed using either of the equalities
\begin{multline}
\bar{r}_{\of{k,l},m}=\sum\limits_{i,j=0}^{N-1}\bar{r}_{k,i}\,\ssof{A^{i}}\indices{^m_j}\,\bar{r}_{l,j}\\
=\sum\limits_{i,j=0}^{N-1}\bar{r}_{l,j}\,\ssof{A^{j}}\indices{^m_i}\,\bar{r}_{k,i}\ ,\label{eq:productrule}
\end{multline}
with $\order{N^2}$ operations, thanks to the sparseness of the companion matrix $A$.

Eq.~\eqref{eq:productrule} also holds in the case where the family of matrix functions is simply given by $f_{k}\of{U}=U^{k}$, $k\in\mathbb{N}$, i.e. by the set of different matrix powers. We then find,
\begin{multline}
a_{k+l,m}=\sum\limits_{i,j=0}^{N-1}\,a_{k,i}\,\ssof{A^i}\indices{^m_j}\,a_{l,j}\\
=\sum\limits_{i,j=0}^{N-1}\,a_{l,j}\,\ssof{A^j}\indices{^m_i}\,a_{k,i}\ ,
\end{multline}
which reduces to the iteration formula \eqref{eq:ankparam} when either $k=1$ (using first equality) or $l=1$ (using second equality), due to \eqref{eq:ankparamc}.

\subsubsection{Shifted power series}\label{sssec:shiftedseries}
There are cases where the function $f\of{x}$ from \eqref{eq:scalfunc} cannot be expanded in a power series around $x=0$. The procedure could then be carried out with a shifted function,
\[
\tilde{f}\of{x}=f\of{x+x_0}\ ,
\] 
and a corresponding matrix $\tilde{U}$ with shifted eigenvalues: 
\[
\tilde{U}=U-\id\,x_0\ ,
\]
where $x_0$ should be chosen so that the expansion coefficients in
\[
\tilde{f}\of{x}=\sum\limits_{n=0}^{\infty}\,\tilde{r}_{n}\,x^{n}\label{eq:ftildeexpanded}
\]
do not diverge. One then finds $f\of{U}$ from
\[
f\of{U}=\tilde{f}\sof{\tilde{U}}=\sum\limits_{i=0}^{N-1}\,\bar{\tilde{r}}_{i}\,\tilde{U}^{i}\ .\label{eq:chfresultshifted}
\]
Of course, this procedure is applicable only if all eigenvalues of $\tilde{U}$ lay within the domain of convergence of \eqref{eq:ftildeexpanded}.

\section{Implementation}\label{sec:implementation}

In the following we describe a possible implementation of the iterative Cayley-Hamilton method for evaluating scalar functions on matrices and for computing the corresponding matrix-valued differentials. An implementation in C++ is available from~\cite{tobias_rindlisbacher_2024_10979006}.

\subsection{Computing function values}\label{ssec:computingfuncvals}
Assume that a matrix $U\in\mathbb{C}^{N\times N}$ and the Taylor series coefficients $r_n$, $n\in\cof{0,1,\ldots,\infty}$ of a function $f\of{x}$ are given. To compute $f\of{U}$ with the iterative Cayley-Hamilton method described in the previous section, we can proceed as follows. As a first step, we need the first $N$ powers of $U$ and their traces. We therefore compute
\begin{subequations}\label{eq:powersofU}
\begin{align}
P_0&=\id\\
P_1&=U\\
 &\vdots\nonumber\\
P_n&=P_{\floor{n/2}}\,P_{\ceil{n/2}}\\
 &\vdots\nonumber\\
P_{N}&=P_{\floor{N/2}}\,P_{\ceil{N/2}}
\end{align}
\end{subequations}
where $\floor{\cdot}$ and $\ceil{\cdot}$ are, respectively, the floor and ceil operations, as well as,
\[
p_n=\trace\of{P_n}\ ,\ \forall n\in\cof{0,\ldots,N}\ .
\]

The $p_n$, $n\in\cof{0,\ldots,N}$ are the first $N+1$ power sums of the $N$ eigenvalues, $\cof{\lambda_0,\ldots,\lambda_{N-1}}$, of $U$, since
\[
p_n=\trace\of{U^n}=\sum_{i=0}^{N-1}\,\lambda_i^{n}\ .
\]

\iffalse
The coefficients, $c_i$, $i\in\cof{0,\ldots,N}$, of the characteristic polynomial of $U$, can now be determined noting that
\[
c_i=\of{-1}^{i+N}\,e_{N-i}\ \forall\,i\,\in\,\cof{0,\ldots,N}\ ,\label{eq:chpolycoeffs}
\]
where the $e_i$, $i\in\cof{0,\ldots,N}$ are the elementary symmetric polynomials defined by 
\[
\prod\limits_{i=0}^{N-1}\of{1+x_i\,t}=\sum\limits_{j=0}^{N}\,e_j\,t^{j}\ ,
\]
and which, using Newton's identities, can be obtained from the Eigenvalue power sums $p_n$, $n\in\cof{0,\ldots,N}$ via forward iteration:
\begin{subequations}
\begin{align}
e_0&=1\\
e_n&=\frac{1}{n}\,\sum\limits_{i=1}^{n}\,\of{-1}^{i+1}\,p_{i}\,e_{n-i}\ ,
\end{align}
\end{subequations}
for $0<n\leq N$.\\
\fi

Next we need the coefficients, $c_i$, $i\in\cof{0,\ldots,N}$, of the characteristic polynomial of $U$, which, using Newton's identities for elementary symmetric polynomials, can now be determined by backward iteration, starting from
\begin{subequations}\label{eq:chpolycoeffs}
\[
c_N=1\ ,
\]
and successively using
\[
c_{N-n}=-\frac{1}{n}\,\sum\limits_{i=1}^{n}\,p_{i}\,c_{N-n+i}\ 
\]
\end{subequations}
for $n=1,\ldots,N$ till $c_0$ is reached. 

Note that the $N$-th power of $U$, $P_{N}=U^N$, is computed in \eqref{eq:powersofU} only because its trace, $p_{N}=\trace\ssof{U^N}$, is needed to compute the characteristic polynomial coefficient $c_0=\of{-1}^{N}\det\of{U}$. If $\det\of{U}$ is known, e.g. because $U$ is unitary, the computation of $P_{N}$ can be skipped and $c_0$ set to the known value. 

Finally, the computation of the coefficients $\cof{\bar{r}_{0},\ldots,\bar{r}_{N-1}}$, with
\[
\bar{r}_{i}=\sum\limits_{n=0}^{\infty}\,r_{n}\,a_{n,i}\ ,\label{eq:rbaricomp}
\]
can be carried out as described in the pseudo-code shown in Alg.~\ref{algo:ch_coeff}. The algorithm takes as input the non-trivial characteristic polynomial coefficients $\cof{c_i}_{i\in\cof{0,\ldots,N-1}}$ and a function $r\of{n}=r_n$ that produces the sequence of power series coefficients $r_n$ from \eqref{eq:scalfunc}. The algorithm stops to produce and sum further terms as soon as the coefficients $\cof{\bar{r}_{0},\ldots,\bar{r}_{N-1}}$ have stopped changing (within machine precision) for nhl\_max consecutive iterations.

Note that the computational cost of each iteration step in Alg.~\ref{algo:ch_coeff} is $\order{N}$.
 
\begin{algorithm}[!h]
\SetArgSty{textrm}
\SetFuncArgSty{textrm}
\SetStartEndCondition{ (}{)}{)}%
%\SetAlgoBlockMarkers{\{}{\}}%
%\SetKwFor{For}{for}{}{}%
%\SetKwFor{ForEach}{foreach}{}{}%
%\SetKwIF{If}{ElseIf}{Else}{if}{}{else if}{else}{}%
%\SetKwFor{While}{while}{}{}%
%\SetKwRepeat{Repeat}{repeat}{until}%
%\AlgoDisplayBlockMarkers\SetAlgoNoLine%
\SetKwFor{For}{for}{ \{}{}%
\SetKwFor{ForEach}{foreach}{}{}%
\SetKwIF{If}{ElseIf}{Else}{if}{ \{}{else if}{else \{}{}%
\SetKwFor{While}{while}{ \{}{}%
\SetKwRepeat{Repeat}{repeat \{}{until}%
\AlgoDisplayBlockMarkers%
\AlgoDisplayGroupMarkers%
\SetAlgoBlockMarkers{}{\}}%
\SetAlgoNoLine\SetAlgoNoEnd%
\SetKwArray{Pal}{a}
\SetKwArray{Rspc}{c}
\SetKwArray{Tal}{rtot}
\SetKwData{Tmp}{trtot}
\SetKwData{Tch}{changed}
\SetKwData{Nhlmax}{nhl\_max}
\SetKwData{Maxit}{maxit}
\SetKwData{Nhl}{nhl}
\SetKwData{Fc}{rn}
\SetKwData{AFc}{afc}
\SetKwData{Ch}{ta}
\SetKwData{Cho}{tao}
\SetKwData{ACh}{ach}
\SetKwData{Rsc}{s}
\SetKwData{Rsci}{si}
\SetKwData{Rsctot}{sr}
\SetKwData{Ii}{i}
\SetKwData{Im}{m}
\SetKwData{Ij}{j}
\SetKwData{Ik}{k}
\SetKwData{In}{n}
\SetKwData{Nmax}{N}
\SetKw{And}{$\&\&$}
\SetKw{Or}{$\|\|$}
\SetKw{Eq}{$==$}
\SetKw{SetTo}{$=$}
\SetKw{Neq}{$\neq$}
\SetKw{Gt}{$>$}
\SetKw{Lt}{$<$}
\SetKw{Geq}{$\geq$}
\SetKw{Leq}{$\leq$}
\SetKw{Mult}{$*$}
\SetKw{Div}{$/$}
\SetKw{Plus}{$+$}
\SetKw{Minus}{$-$}
\SetKw{PlusEq}{$+\!=$}
\SetKw{MinusEq}{$-\!=$}
\SetKw{Incr}{\!$++$}
\SetKw{Decr}{\!$--$}
\SetKw{Break}{break}
\SetKw{Itype}{itype}
\SetKw{Ctype}{ctype}
\SetKw{Ftype}{ftype}
\SetKwFunction{R}{r}
\SetKwFunction{Abs}{abs}
\SetKwFunction{Abssq}{abssq}
\SetKwFunction{Sqrt}{sqrt}
\SetKwInOut{Input}{input}
\SetKwInOut{Output}{output}
\SetKwInOut{Local}{local}
\Input{matrix size \Nmax\;\\
array \Rspc$=\cof{c_{0},\ldots,c_{N-1}}$\;\\
function \R{\In}$=r_n$\;\\
stop criterion: \Nhlmax=\ 3\;\\
max. iterations: \Maxit=\ 10\ \Mult\Nmax\;}
\Output{array \Tal$=\cof{\bar{r}_{0},\ldots,\bar{r}_{N-1}}$\;}
\Local{temporary array of size \Nmax: \Pal\;\\
temporary \Itype vars: \In,\ \Ik,\ \Nhl,\ \Tch\;\\
temporary \Ctype vars: \Tmp,\ \Cho\;\\
{\color{gray}temporary \Ftype vars: \Rsc,\ \Rsci,\ \Rsctot\;}
}
\BlankLine
\For{\Ik\SetTo 0; \Ik\Lt\Nmax; \Ik\Incr}{
  \Tal{\Ik}\ \SetTo \R{\Ik}\;
  \Pal{\Ik}\ \SetTo 0\;
}
\Pal{\Nmax\Minus 1}\ \SetTo 1.0\;
\Nhl\SetTo \Nhlmax\;
{\color{gray}\Rsctot\SetTo 1.0;\ \ \Rsci\SetTo 1.0\;}
\For{\In\SetTo \Nmax; \In\Lt \Maxit; \In\Incr}{
  {\color{gray}\Rsc\SetTo 0\;}
  \Fc\SetTo\R{\In}\ {\color{gray}\Mult\Rsctot}\;
  \Tch\SetTo 0\;
  \Cho\SetTo \Pal{\Nmax\Minus 1}\ \Mult\Rsci\;
  \For{\Ik\SetTo \Nmax\Minus 1; \Ik\Gt 0; \Ik\Decr}{
    \Pal{\Ik}\ \SetTo\Pal{\Ik\Minus 1}\ \Mult\Rsci\ \Minus\ \Cho\ \Mult\Rspc{\Ik}\;
    {\color{gray}\Rsc\PlusEq\Abssq{\Pal{\Ik}}\;}
    \Tmp\SetTo\Tal{\Ik}\ \Plus \Fc\Mult\Pal{\Ik}\;
    \If{\Tmp\Neq\Tal{\Ik}}{
      \Tal{\Ik}\ \SetTo \Tmp\;
      \Tch\SetTo 1\;
    }
  }
  \Pal{0}\ \SetTo\ \Minus\Cho\ \Mult\Rspc{0}\;
  {\color{gray}\Rsc\PlusEq\Abssq{\Pal{0}}\;}
  \Tmp\SetTo\Tal{0}\ \Plus \Fc\Mult\Pal{0}\;
  \If{\Tmp\Neq\Tal{0}}{
    \Tal{0}\ \SetTo \Tmp\;
    \Tch\SetTo 1\;
  }
  \lIf*{\Tch\Gt 0}{\\  
    \hskip1.8em\Nhl\SetTo\Nhlmax;\\
  }
  \lElse*{\\
     \hskip1.8em\Nhl\Decr;\ \ 
     \lIf*{\Nhl\Leq 0}{
       \Break;\ 
     }\\
  }\\
  {\color{gray}
  \lIf*{\Rsc\Gt 1.0}{\\
     \hskip1.8em\Rsc\SetTo \Sqrt{\Rsc};\ \ \Rsctot\SetTo\Rsctot\ \Mult\Rsc;\ \ \Rsci\SetTo 1.0\ \Div\Rsc;\\
  }
  \lElse*{\\
     \hskip1.8em\Rsci\SetTo 1.0;\\
  }\\
  }
}
\caption{Computation of the coefficients $\cof{\overline{r}_{0},\ldots,\overline{r}_{N-1}}$ for \eqref{eq:chfresult}, using the characteristic polynomial coefficients, $\cof{c_{0},\ldots,c_{N-1}}$ from \eqref{eq:chpolycoeffs} and the coefficients $r_{n}$ from \eqref{eq:scalfunc}. The gray pieces of code are added for stability, as explained in Sec.~\ref{sssec:stabiteration}.}\label{algo:ch_coeff}
\end{algorithm}

\subsection{Coping with round-off errors}\label{ssec:roundofferrors}
Since the computation of the characteristic polynomial coefficients according to \eqref{eq:chpolycoeffs} involves summation of terms of potentially opposite signs, there is a reasonable possibility that the resulting coefficient values get affected by round-off errors. This can in particular happen if the summed floating point numbers are very different in magnitude. And of course, this problem might also occur in the iterative computation of the coefficients $\cof{a_{n,i}}_{n\in\mathbb{N}_0,i\in\cof{0,\ldots,N-1}}$ according to \eqref{eq:ankparam}. 

In the following, we discuss two counter measures that can be taken to avoid running into problems due to round-off errors.  

\subsubsection{Re-scaling of input matrix}\label{sssec:rescaledinputmat}
We note that \eqref{eq:monodecomp} implies that the Cayley-Hamilton coefficients $\cof{a_{n,i}}_{n\in\mathbb{N}_0,i\in\cof{0,\ldots,N-1}}$ for a matrix $U$ can be related as follows to the corresponding coefficients for a scaled matrix $\bar{U}=s\,U$:
\[
a_{n,i}\ssof{\bar{U}}=a_{n,i}\of{U}\,s^{n-i}\ .\label{eq:rescledchcoeffs}
\]
And since $a_{N,i}=-c_{i}$ for $i\in\cof{0,\ldots,N-1}$, one has for the characteristic polynomial coefficients:
\[
c_{i}\ssof{\bar{U}}=c_{i}\of{U}\,s^{N-i}\ .
\]
This means, that we can always perform the computation of the characteristic polynomial coefficients and of the $\cof{a_{n,i}}_{n\in\mathbb{N}_0,i\in\cof{0,\ldots,N-1}}$ by using an appropriately scaled matrix $\bar{U}$. The so obtained Cayley-Hamilton coefficients, $\cof{a_{n,i}\ssof{\bar{U}}}_{n\in\mathbb{N}_0,i\in\cof{0,\ldots,N-1}}$, can then be re-scaled when computing the coefficients $\cof{\bar{r}_{i}}_{i\in\cof{0,\ldots,N-1}}$, using
\[
\bar{r}_{i}=\sum\limits_{n=0}^{\infty}\,r_{n}\,a_{n,i}\ssof{\bar{U}}\,s^{i-n}\ ,\label{eq:rbaricomprescaled}
\]
instead of \eqref{eq:rbaricomp}. Note that the $\bar{r}_{i}$ obtained from \eqref{eq:rbaricomprescaled} are the same as those obtained via \eqref{eq:rbaricomp} from the original, unscaled matrix $U$. One therefore still has
\[
f\of{U}=\sum\limits_{i=0}^{N-1}\,\bar{r}_i\,P_i\ ,\label{eq:fresrbariscaled}
\]
with the $P_i=U^i$ $i\in\cof{0,\ldots,N}$ being the matrix powers of the original matrix $U$.
When working instead with the powers $\tilde{P}_i=\bar{U}^i$, $i\in\cof{0,\ldots,N}$ of the scaled matrix $\bar{U}=s\,U$, the factor $s^i$ in \eqref{eq:rbaricomprescaled} can be dropped, i.e. the coefficients then read   
\[
\tilde{\bar{r}}_{i}=\sum\limits_{n=0}^{\infty}\,r_{n}\,a_{n,i}\ssof{\bar{U}}\,s^{-n}\ ,\label{eq:rbaricomprescaledr}
\]
and $f\of{U}$ is computed as
\[
f\of{U}=\sum\limits_{i=0}^{N-1}\,\tilde{\bar{r}}_i\,\tilde{P}_i\ .\label{eq:fresrbariscaledr}
\]
\iffalse
If the scaling factor $s$ is known/computed in advance, then \eqref{eq:rbaricomprescaledr} and \eqref{eq:fresrbariscaledr} are presumably preferable. To determine an appropriate scaling factor, the matrix powers, $P_i=U^i$, $i\in\cof{0,\ldots,N}$, of the unscaled matrix $U$ might, however, be useful and therefore computed first. One can then proceed with the computation of the Cayley-Hamilton coefficients, by scaling only the traces, $p_i=\trace\ssof{U^i}$, $i\in\cof{1,\ldots,N}$, of the powers of $U$, and compute $f\of{U}$ from \eqref{eq:fresrbariscaled}, using \eqref{eq:rbaricomprescaled} and the unscaled matrix powers $P_i=U^i$, $i\in\cof{0,\ldots,N}$. 
\fi

\subsubsection{Stabilizing the Cayley-Hamilton iteration}\label{sssec:stabiteration}
Regardless of whether the input matrix $U$ has been re-scaled or not, the iterative computation of the Cayley-Hamilton coefficients $\cof{a_{n,i}\of{U}}_{n\in\mathbb{N}_0,i\in\cof{0,\ldots,N-1}}$ can be stabilized by normalizing the coefficients $a_{n-1,i}\of{U}$, $i\in\cof{0,\ldots,N-1}$  before they are used to compute the coefficients $a_{n,i}\of{U}$, $i\in\cof{0,\ldots,N-1}$.  

To justify this a bit better, we return to the representation of the Cayley-Hamilton iteration from Eq.~\eqref{eq:vectorchiter}, and note that the companion matrix \eqref{eq:companionmat} has the same eigenvalues as the matrix $U$. Hence, the norms of the vectors $v_{n}$, $n\in\mathbb{N}_0$, in \eqref{eq:vectorchiter} will for large $n$ scale like the $n$-th power of the magnitude of the largest eigenvalue of $U$. Normalizing the vector $v_{n-1}$ before applying $A$ to it to compute $v_n$ prevents  problems arising from the magnitudes of the $v_{n}$, $n\in\mathbb{N}_0$ growing too large.

To be more precise, we can write each vector $v_{n}$ as
\[
v_{n}=\alpha_n\,\hat{v}_n\ ,
\] 
with $\alpha_n=\ssabs{v_{n}}$ and $\hat{v}_n=v_n/\alpha_n$. In terms of the pairs $\ssof{\alpha_n,\hat{v}_n}$, $n\in\mathbb{N}_0$, the iteration rule \eqref{eq:vectorchiter} would then take the following form:
\begin{subequations}\label{eq:vectorchiterregt}
\begin{align}
\tilde{v}_n &=A\,\hat{v}_{n-1}\ ,\\
\ssof{\alpha_n,\hat{v}_n} &=\ssof{\abs{\tilde{v}_n}\,\alpha_{n-1},\tilde{v}_n/\abs{\tilde{v}_n}}\ ,
\end{align} 
\end{subequations}
for all $n\in\mathbb{N}$. The normalization of $\tilde{v}_n$ should, however, be performed only if $\ssabs{\tilde{v}_n}>1$, in order to avoid amplifying round-off errors. We therefore modify \eqref{eq:vectorchiterregt} slightly, and use instead:
\begin{subequations}\label{eq:vectorchiterreg}
\begin{align}
\tilde{v}_n &=A\,\hat{v}_{n-1}\ ,\\
\tilde{\alpha}_n &=\max\of{1,\ssabs{\tilde{v}_n}}\ ,\label{eq:vectorchiterregb}\\
\ssof{\alpha_n,\hat{v}_n} &=\ssof{\tilde{\alpha}_n\,\alpha_{n-1},\tilde{v}_n/\tilde{\alpha}_n}\ .
\end{align} 
\end{subequations}
In Alg.~\ref{algo:ch_coeff} the pieces of code required to implement the just described procedure are displayed in gray.  

It is also worth noting, that the stabilization procedure described in this section is in principle only necessary if the magnitude of the largest eigenvalue of the input matrix $U$ (and therefore of $A$) is larger than $1$; if it is smaller than $1$ one has
\[
A^n\ \stackrel{\of{n\to\infty}}{\longrightarrow}\ 0\ ,\label{eq:convmatpow}
\]
and therefore
\[
v_n=A\,v_{n-1}=A^n\,v_0\ \stackrel{\of{n\to\infty}}{\longrightarrow}\ 0\ .\label{eq:convmatpowvec}
\]
The step \eqref{eq:vectorchiterregb} ensures that \eqref{eq:convmatpowvec} is in this case also true for the sequence of vectors $\hat{v}_n$, $n=0,1,2,\ldots$, produced by the iteration procedure \eqref{eq:vectorchiterreg}.

\subsection{Computing derivative terms}\label{ssec:computingderivterms}
In practical applications it is often necessary to compute not only $f\of{U}$ but also the corresponding matrix valued differentials, 
\[
\dd f\of{U}=\partd{f\of{U}}{U\indices{^a_b}}\,\dd  U\indices{^a_b}\ .\label{eq:fdifferential}
\]
There are two different approaches to compute these quantities with the Cayley-Hamilton method, which differ in whether the Cayley-Hamilton expansion or the differentiation operation is first applied to $f\of{U}$.

\subsubsection{Differentail of Cayley-Hamilton expansion}\label{sssec:chainrule}
Plugging in the Cayley-Hamilton expansion of $f\of{U}$ from \eqref{eq:chfresult}, the differential \eqref{eq:fdifferential} can be written as
\[
\dd f\of{U}=\sum\limits_{i=0}^{N-1}\of{P_i\,\dd\bar{r}_{i}+\bar{r}_{i}\,\dd P_i}\ ,\label{eq:chfdiff}
\]
with $P_i$ $i\in\cof{0,\ldots,N}$ being the matrix powers of $U$ given in \eqref{eq:powersofU}.
To evaluate \eqref{eq:chfdiff}, we need to compute the differentials of the $P_i$ and $\bar{r}_i$ for $i\in\cof{0,\ldots,N-1}$, where for the latter, we will in general also need the differential of $P_N$. 

The differentials of the $P_i$, $i\in\cof{0,\ldots,N}$ are easily computed by iteration, using that for $i\geq 2$, one has
\begin{subequations}\label{eq:dpowersofU}
\[
\dd P_i=U\,\dd P_{i-1}+\of{\dd P_{i-1}}\,U-U\,\of{\dd P_{i-2}}\,U\ ,
\]
with
\[
\dd P_0=0\quad,\quad \dd P_1=\dd U\quad .
\]
\end{subequations}
Correspondingly, the differentials of the traced powers of $U$, $p_i=\trace\of{P_i}$, $i\in\cof{0,\ldots,N}$, are given by 
\[
\dd p_i=\trace\of{\dd P_i}\quad,\quad i\in\cof{0,\ldots,N}\ .\label{eq:dtrpowersofU}
\]

Note that in certain cases one might only be interested in the $\cof{\dd \bar{r}_i}_{i\in\cof{0,\ldots,N-1}}$ in \eqref{eq:chfdiff} (cf. Sec.~\ref{ssec:onelinkint}), in which case the $\cof{\dd P_i}_{i\in\cof{0,\ldots,N}}$ themselves are not required and the expensive computation in \eqref{eq:dpowersofU} can be skipped by computing \eqref{eq:dtrpowersofU} directly as
%\[
%\dd p_i = \partd{p_i}{U\indices{^b_a}}\,\dd U\indices{^b_a} = i\,\of{P_{i-1}}\indices{^a_b}\,\dd U\indices{^b_a} \ .
%\] 
\begin{subequations}\label{eq:dtrp}
\begin{align}
\of{\dd p_0}\indices{^a_b} &= \partd{p_0}{U\indices{^b_a}} = 0\ ,\\
\of{\dd p_i}\indices{^a_b} &= \partd{p_i}{U\indices{^b_a}} = i\,\of{P_{i-1}}\indices{^a_b}\quad\forall i\in\cof{1,\ldots,N}\ .
\end{align}
\end{subequations}

Next we need to compute the differentials of the characteristic polynomial coefficients, $c_i$, $i\in\cof{0,\ldots,N}$ of $U$. From \eqref{eq:chpolycoeffs}, these could be obtained by setting 
\begin{subequations}\label{eq:dchpolycoeffs}
\[
\dd c_N=0\ ,
\]
and successively
\begin{multline}
\dd c_{N-n}=-\frac{1}{n}\,\sum\limits_{i=1}^{n}\,\of{c_{N-n+i}\,\dd p_{i}+p_{i}\,\dd c_{N-n+i}}\ ,
\end{multline}
\end{subequations}
for $n=1,\ldots,N$ till the expression for $\dd c_0$ is reached. By expanding this recurrence relation one arrives at the following explicit expressions:
\[
\dd c_{N-n}=-\sum\limits_{i=1}^{n}\frac{1}{i}\,c_{N-n+i}\,\dd p_{i} \quad \forall n\in\cof{0,\ldots,N}\ .\label{eq:charpolycoeffdiff}
\]

For the differentials of the Cayley-Hamilton coefficients $\cof{a_{n,i}}_{n\in\mathbb{N}_0,i\in\cof{0,\ldots,N-1}}$ one finds from \eqref{eq:ankparam} for $n\geq N$, $0< k\leq N-1$ the recurrence relations
\begin{subequations}\label{eq:dankparam}
\begin{align}
\dd a_{n,0}&=-\dd a_{n-1,N-1}\,c_{0}-a_{n-1,N-1}\,\dd c_{0}\ ,\\
\dd a_{n,k}&=\dd a_{n-1,k-1}-\dd a_{n-1,N-1}\,c_{k}\nonumber\\
&\hskip9em -a_{n-1,N-1}\,\dd c_{k}\ ,
\end{align}
with initial conditions
\[
\dd a_{n,k}=0\ ,\ 0\leq n<N\ ,\ 0\leq k\leq N-1\ .
\]
\end{subequations}

Using the latter, the differentials of the expansion coefficients $\bar{r}_i$, $i\in\cof{0,\ldots,N-1}$ are obtained as:
\[
\dd\bar{r}_{i}=\sum\limits_{n=0}^{\infty}\,r_{n}\,\dd a_{n,i}\ .\label{eq:diffcoeffs}
\]
\iffalse
The pseudo code in Alg.~\ref{algo:ch_coeff_incl_diff} illustrates how the computation of the differnetials \eqref{eq:diffcoeffs} can be incorporated in the algorithm for computing the $\cof{\bar{r}_{i}}_{i\in\cof{0,\ldots,N-1}}$, shown in Alg.~\ref{algo:ch_coeff}. For brevity, Alg.~\ref{algo:ch_coeff_incl_diff} merely shows how the
\[
\dd\bar{r}_{i}=\partd{\bar{r}_i}{U\indices{^a_b}}\dd U\indices{^a_b}\quad,\quad i\in\cof{0,\ldots,N-1}
\]
are computed for a particular set of differentials of the characteristic polynomial coefficients 
\[
\dd c_{i}=\partd{c_i}{U\indices{^a_b}}\dd U\indices{^a_b}\quad,\quad i\in\cof{0,\ldots,N-1}\ ,
\]
corresponding to a particular choice of $\dd U$. The computation can, however, easily be extended to run over a complete basis of the co-tangent space of $\mathbb{C}^{n\times n}$.
\fi
Unfortunately, the here described method for computing $\dd f\of{U}$ is with regard to computational cost a bit on the expensive side, since the computation of the differentials of the matrix powers, $P_i$, $i\in\cof{0,\ldots,N}$ requires many matrix multiplications. We therefore discuss next a more economic method for obtaining the coefficients of $\dd f\of{U}$ in \eqref{eq:fdifferential}.  
\subsubsection{Cayley-Hamilton expansion of differential}\label{sssec:chexpansionofderivtaylorseries}
We start by plugging into \eqref{eq:fdifferential} the expression for $f\of{U}$ in terms of its defining power series:
\[
f\of{U}=\sum\limits_{n=0}^{\infty}\,r_{n}\,U^{n}\ .
\]
The components of the partial derivative term in \eqref{eq:fdifferential} are then given by:
\begin{multline}
\partd{f\of{U}\indices{^a_b}}{U\indices{^c_d}}=\sum\limits_{n=0}^{\infty}\,r_{n}\,\partd{\of{U^{n}}\indices{^a_b}}{U\indices{^c_d}}\\
= \sum\limits_{n=0}^{\infty}\,r_{n+1} \sum\limits_{m=0}^{n} \of{U^{m}}\indices{^a_c} \of{U^{n-m}}\indices{^d_b}\ .\label{eq:chfdifftaylor}
\end{multline}
We now apply the Cayley-Hamilton expansion to the matrix powers $U^m$ and $U^{n-m}$ in \eqref{eq:chfdifftaylor} to obtain:
\begin{multline}
\partd{f\of{U}\indices{^a_b}}{U\indices{^c_d}}=\\
\sum\limits_{n=0}^{\infty}\sum\limits_{i,j=0}^{N-1}r_{n+1}\bof{\underbrace{\sum\limits_{m=0}^{n} a_{m,i}\,a_{n-m,j}}_{a_{n,i,j}}}\,\of{U^i}\indices{^a_c} \of{U^j}\indices{^d_b}\\
=\sum\limits_{i,j=0}^{N-1}\,\bof{\underbrace{\sum\limits_{n=0}^{\infty} r_{n+1}\,a_{n,i,j}}_{\overline{r}_{i,j}}}\,\of{U^i}\indices{^a_c} \of{U^j}\indices{^d_b}\\
=\sum\limits_{i,j=0}^{N-1} \overline{r}_{i,j}\,\of{U^i}\indices{^a_c} \of{U^j}\indices{^d_b}\ .\label{eq:rmatcoeffdef}
\end{multline}
Note that the $a_{n,i,j}$ for $n\in\cof{0,1,\ldots}$ are symmetric in the indices $i,j\,\in\,\cof{0,\ldots,N-1}$.
The coefficients $\cof{a_{n,i,j}}_{i,j\in\cof{0,\ldots,N-1},n\in\cof{0,1,\ldots}}$ can be computed iteratively. To see this, we write out the expression for $a_{n+1,i,j}$:
\begin{multline}
a_{n+1,i,j} = \sum\limits_{m=0}^{n+1} a_{m,i}\,a_{n+1-m,j}\\
= \sum\limits_{m=0}^{n} a_{m,i}\,\underbrace{a_{n+1-m,j}}_{\eqref{eq:ankparam}} + a_{n+1,i}\,\underbrace{a_{0,j}}_{\delta_{0,j}}\ ,
\end{multline}
where on the second line, it is indicated that the iteration procedure for the Cayley-Hamilton coefficients from \eqref{eq:ankparam} can be used to re-express the summed terms. Doing so leads to the following recurrence relation for the $a_{n,i,j}$: 
\begin{subequations}\label{eq:dkmatparam}
\begin{align}
a_{n+1,i,0}&=a_{n+1,i} - a_{n,i,N-1}\,c_0\\
a_{n+1,i,j}&=a_{n,i,j-1} - a_{n,i,N-1}\,c_j\ ,
\end{align}
with the initial conditions
\[
a_{N-1,i,j}=\sum_{m=0}^{N-1}\,\delta_{i,m}\,\delta_{N-1-m,j}\ .
\]
\end{subequations}

\begin{figure*}
\begin{minipage}[t]{0.49\textwidth}
\begin{algorithm}[H]
\ContinuedFloat*
\SetArgSty{textrm}
\SetFuncArgSty{textrm}
\SetStartEndCondition{ (}{)}{)}%
%\SetAlgoBlockMarkers{\{}{\}}%
%\SetKwFor{For}{for}{}{}%
%\SetKwFor{ForEach}{foreach}{}{}%
%\SetKwIF{If}{ElseIf}{Else}{if}{}{else if}{else}{}%
%\SetKwFor{While}{while}{}{}%
%\SetKwRepeat{Repeat}{repeat}{until}%
%\AlgoDisplayBlockMarkers\SetAlgoNoLine%
\SetKwBlock{Blockb}{\{}{}
\SetKwBlock{Blocke}{}{\}}
\SetKwFor{For}{for}{ \{}{\}}%
\SetKwFor{Forb}{for}{ \{}{}%
\SetKwBlock{Fore}{}{\}}
\SetKwIF{If}{ElseIf}{Else}{if}{ \{}{\} else if}{\} else \{}{\}}%
\SetKwFor{While}{while}{ \{}{\}}%
\AlgoDisplayBlockMarkers%
\AlgoDisplayGroupMarkers%
\SetAlgoBlockMarkers{}{}%
\SetAlgoNoLine%
%\SetAlgoNoEnd%
\SetKwArray{Pal}{a}
\SetKwArray{Rspc}{c}
\SetKwArray{Tal}{rtot}
\SetKwArray{Tmal}{rmtot}
\SetKwArray{Palk}{am}
\SetKwData{Tmp}{trtot}
\SetKwData{Nhlmax}{nhl\_max}
\SetKwData{Maxit}{maxit}
\SetKwData{Nhl}{nhl}
\SetKwData{Fc}{rn}
\SetKwData{TotFc}{totrn}
\SetKwData{TmpFc}{trn}
\SetKwData{Cho}{tao}
\SetKwData{Rsc}{s}
\SetKwData{Rsci}{si}
\SetKwData{Rsctot}{sr}
\SetKwData{Rsctotr}{srr}
\SetKwData{Ii}{i}
\SetKwData{Im}{m}
\SetKwData{Ij}{j}
\SetKwData{Ik}{k}
\SetKwData{In}{n}
\SetKwData{Il}{l}
\SetKwData{Nmax}{N}
\SetKw{And}{$\&\&$}
\SetKw{Or}{$\|\|$}
\SetKw{Eq}{$==$}
\SetKw{SetTo}{$=$}
\SetKw{Neq}{$\neq$}
\SetKw{Gt}{$>$}
\SetKw{Lt}{$<$}
\SetKw{Geq}{$\geq$}
\SetKw{Leq}{$\leq$}
\SetKw{Mult}{$*$}
\SetKw{Div}{$/$}
\SetKw{Plus}{$+$}
\SetKw{Minus}{$-$}
\SetKw{PlusEq}{$+\!=$}
\SetKw{MinusEq}{$-\!=$}
\SetKw{Incr}{\!$++$}
\SetKw{Decr}{\!$--$}
\SetKw{Break}{break}
\SetKw{Itype}{itype}
\SetKw{Ctype}{ctype}
\SetKw{Ftype}{ftype}
\SetKwFunction{R}{r}
\SetKwFunction{Abs}{abs}
\SetKwFunction{Abssq}{abssq}
\SetKwFunction{Sqrt}{sqrt}
\SetKwInOut{Input}{input}
\SetKwInOut{Output}{output}
\SetKwInOut{Local}{local}
\LinesNumbered
\Input{matrix size \Nmax\;\\
array \Rspc$=\cof{c_{0},\ldots,c_{N-1}}$\;\\
function \R{\In}$=r_n$\;\\
stop criterion: \Nhlmax=\ 3\;\\
max. iterations: \Maxit=\ 10\ \Mult\Nmax\;}
\Output{array \Tal$=\cof{\bar{r}_{0},\ldots,\bar{r}_{N-1}}$\;\\
matrix \Tmal$=\cof{\bar{r}_{i,j}}_{i,j\in\cof{0,\ldots,N-1}}$\;}
\BlankLine
\For{\Ik\SetTo 0; \Ik\Lt\Nmax; \Ik\Incr}{
  \Pal{\Ik}\ \SetTo 0\;
  \For{\Ij\SetTo \Ik; \Ij\Lt\Nmax; \Ij\Incr}{
    \Tmal{\Ik,\Ij}\ \SetTo 0\;
    \Palk{\Ik,\Ij}\ \SetTo 0\;
  }
}
\Pal{\Nmax\Minus 1}\ \SetTo 1.0\;
%\Pal{}\ \SetTo 0;\ \Tmal{}\ \SetTo 0;\ \Palk{}\ \SetTo 0\;
\Fc\SetTo\R{0}\;
\TotFc\SetTo \Fc\;
\For{\Ik\SetTo 0; \Ik\Lt\Nmax; \Ik\Incr}{
  \Tal{\Ik}\ \SetTo \Fc\;
  \Fc\SetTo\R{\Ik\Plus 1}\;
  \TotFc\PlusEq \Fc\;
  \Il\SetTo \Ik\Div 2\;
  \For{\Ij\SetTo \Ik\Minus \Il; \Ij\Leq\Ik; \Ij\Incr}{
    \Tmal{\Ik\Minus \Ij, \Ij}\ \SetTo \Fc\;
  }
}
\Il\SetTo (\Nmax\Minus 1)\Div 2\;
\For{\Ij\SetTo \Nmax\Minus 1\Minus \Il; \Ij\Lt\Nmax; \Ij\Incr}{
  \Palk{\Nmax\Minus 1\Minus \Ij, \Ij}\ \SetTo 1.0\;
}

\Nhl\SetTo 0\;
\Rsctot\SetTo 1.0;\ \Rsctotr\SetTo 1.0;\ \Rsci\SetTo 1.0\;
\Forb{\In\SetTo \Nmax; \In\Lt \Maxit; \In\Incr}{
  \Rsc\SetTo 0\;
  \Cho\SetTo \Pal{\Nmax\Minus 1}\ \Mult\Rsci\;
  \For{\Ik\SetTo \Nmax\Minus 1; \Ik\Gt 0; \Ik\Decr}{
    \Pal{\Ik}\ \SetTo\Pal{\Ik\Minus 1}\ \Mult\Rsci\ \Minus\ \Cho\ \Mult\Rspc{\Ik}\;
    \Rsc\PlusEq\Abssq{\Pal{\Ik}}\;
    \Tal{\Ik}\ \PlusEq\Fc\Mult\Pal{\Ik}\;
  }
  \Pal{0}\ \SetTo\ \Minus\Cho\ \Mult\Rspc{0}\;
  \Rsc\PlusEq\Abssq{\Pal{0}}\;
  \Tal{0}\ \PlusEq\Fc\Mult\Pal{0}\;
}
\end{algorithm}
\end{minipage}\hfill
\begin{minipage}[t]{0.49\textwidth}
\begin{algorithm}[H]
\ContinuedFloat
\SetArgSty{textrm}
\SetFuncArgSty{textrm}
\SetStartEndCondition{ (}{)}{)}%
%\SetAlgoBlockMarkers{\{}{\}}%
%\SetKwFor{For}{for}{}{}%
%\SetKwFor{ForEach}{foreach}{}{}%
%\SetKwIF{If}{ElseIf}{Else}{if}{}{else if}{else}{}%
%\SetKwFor{While}{while}{}{}%
%\SetKwRepeat{Repeat}{repeat}{until}%
%\AlgoDisplayBlockMarkers\SetAlgoNoLine%
\SetKwBlock{Blockb}{\{}{}
\SetKwBlock{Blocke}{}{\}}
\SetKwFor{For}{for}{ \{}{\}}%
\SetKwFor{Forb}{for}{ \{}{}%
\SetKwBlock{Fore}{}{\}}
\SetKwIF{If}{ElseIf}{Else}{if}{ \{}{\} else if}{\} else \{}{\}}%
\SetKwFor{While}{while}{ \{}{\}}%
\AlgoDisplayBlockMarkers%
\AlgoDisplayGroupMarkers%
\SetAlgoBlockMarkers{}{}%
\SetAlgoNoLine%
%\SetAlgoNoEnd%
\SetKwArray{Pal}{a}
\SetKwArray{Rspc}{c}
\SetKwArray{Tal}{rtot}
\SetKwArray{Tmal}{rmtot}
\SetKwArray{Palk}{am}
\SetKwArray{Chko}{tamo}
\SetKwData{Tmp}{trtot}
\SetKwData{Tch}{changed}
\SetKwData{Nhlmax}{nhl\_max}
\SetKwData{Maxit}{maxit}
\SetKwData{Nhl}{nhl}
\SetKwData{Fc}{rn}
\SetKwData{TotFc}{totrn}
\SetKwData{TmpFc}{trn}
\SetKwData{Cho}{tao}
\SetKwData{Rsc}{s}
\SetKwData{Rsci}{si}
\SetKwData{Rsctot}{sr}
\SetKwData{Rsctotr}{srr}
\SetKwData{Ii}{i}
\SetKwData{Im}{m}
\SetKwData{Ij}{j}
\SetKwData{Ik}{k}
\SetKwData{In}{n}
\SetKwData{Il}{l}
\SetKwData{Nmax}{N}
\SetKw{And}{$\&\&$}
\SetKw{Or}{$\|\|$}
\SetKw{Eq}{$==$}
\SetKw{SetTo}{$=$}
\SetKw{Neq}{$\neq$}
\SetKw{Gt}{$>$}
\SetKw{Lt}{$<$}
\SetKw{Geq}{$\geq$}
\SetKw{Leq}{$\leq$}
\SetKw{Mult}{$*$}
\SetKw{Div}{$/$}
\SetKw{Plus}{$+$}
\SetKw{Minus}{$-$}
\SetKw{PlusEq}{$+\!=$}
\SetKw{MinusEq}{$-\!=$}
\SetKw{MultEq}{$*\!=$}
\SetKw{DivEq}{$/\!=$}
\SetKw{Incr}{\!$++$}
\SetKw{Decr}{\!$--$}
\SetKw{Break}{break}
\SetKw{Itype}{itype}
\SetKw{Ctype}{ctype}
\SetKw{Ftype}{ftype}
\SetKwFunction{R}{r}
\SetKwFunction{Abs}{abs}
\SetKwFunction{Abssq}{abssq}
\SetKwFunction{Sqrt}{sqrt}
\SetKwInOut{Input}{input}
\SetKwInOut{Output}{output}
\SetKwInOut{Local}{local}
\LinesNumbered
\setcounter{AlgoLine}{36}
\Local{temporary \Itype vars: \In,\ \Ik,\ \Il,\ \Nhl\;\\
temporary \Ctype vars: \Tmp,\ \Cho\;\\
temporary \Ftype vars: \Rsc,\ \Rsci,\ \Rsctot,\ \Rsctotr,\\
\phantom{temporary \Ftype vars:}\ \ \Fc,\ \TotFc,\ \TmpFc\;\\
temporary arrays of size \Nmax: \Pal\;\\
temporary matrix of size \Nmax\!\!$\times$\Nmax: \Palk\;}
\BlankLine
\BlankLine
\BlankLine
\BlankLine
\nonl\Fore{
  \Rsc\SetTo \Sqrt{\Rsc}\;
  \eIf{\Rsc\Gt 1.0}{
    \Rsctot\MultEq\Rsc\;
    \Rsctotr\SetTo 1.0\;
    \Rsci\SetTo 1.0\ \Div\Rsc\;
  }{
    \Rsctotr\SetTo \Rsc\;
    \Rsci\SetTo 1.0\;
  }
  \Fc\SetTo\R{\In\Plus 1}\ \Mult\Rsctot\;
  \For{\Ik\SetTo \Nmax\Minus 1; \Ik\Geq 0; \Ik\Decr}{
    \Cho\SetTo \Palk{\Ik,\Nmax\Minus 1}\ \Mult\Rsci\;
    \For{\Ij\SetTo \Nmax\Minus 1; \Ij\Gt\Ik; \Ij\Decr}{
       \Palk{\Ik,\Ij}\ \SetTo \Palk{\Ik,\Ij\Minus 1}\ \Mult\Rsci\ \Minus \Cho\ \Mult\Rspc{\Ij}\;
       \Tmal{\Ik,\Ij}\ \PlusEq \Fc\Mult \Palk{\Ik,\Ij}\;
    }
    \eIf{\Ik\Gt 0}{
       \Palk{\Ik,\Ik}\ \SetTo \Palk{\Ik\Minus 1,\Ik}\ \Mult\Rsci\ \Minus \Cho\ \Mult\Rspc{\Ik}\;
    }{
       \Palk{\Ik,\Ik}\ \SetTo \Pal{\Ik}\ \Mult\Rsci\ \Minus \Cho\ \Mult\Rspc{\Ik}\;
    }
    \Tmal{\Ik,\Ik}\ \PlusEq \Fc\Mult \Palk{\Ik,\Ik}\;
  }
  \TmpFc\SetTo \TotFc\;
  \TotFc\PlusEq\Fc\Mult\Rsctotr\;
  \eIf{\TotFc\Eq \TmpFc}{
    \Nhl\Incr\;
    \If{\Nhl\Geq\Nhlmax}{
      \Break\;
    }
  }{
    \Nhl\SetTo 0\;
  } 
  
}

\end{algorithm}
\end{minipage}
\begin{algorithm}[H]
\ContinuedFloat
\caption{Computation of the coefficients $\cof{\overline{r}_{0},\ldots,\overline{r}_{N-1}}$ for \eqref{eq:chfresult} and $\cof{\overline{r}_{i,j}}_{i,j\in\cof{0,\ldots,N-1}}$ for \eqref{eq:rmatcoeffdef}, using the characteristic polynomial coefficients, $\cof{c_{0},\ldots,c_{N-1}}$ from \eqref{eq:chpolycoeffs} and the coefficients $r_{n}$ from \eqref{eq:scalfunc}. The code uses the rescaling procedure discussed in Sec.~\ref{sssec:stabiteration} for improving numerical stability and an alternative, simpler convergence criterion, which assumes that the Cayley-Hamilton coefficients have converged if the running sum of the rescaled power series coefficients has converged. On lines 1-25 variables are initialized. Lines 27-36 represent an iteration step \eqref{eq:ankparam} in the rescaled form \eqref{eq:vectorchiterregb} (with $\mathrm{s}\to\tilde{\alpha}_{n+1}$, $\mathrm{si}\to 1/\tilde{\alpha}_{n+1}$, $\mathrm{sr}\to\alpha_{n+1}$), and accumulation of terms for the $\cof{\overline{r}_{0},\ldots,\overline{r}_{N-1}}$ from \eqref{eq:chfresult}. Lines 37-45 determine the rescaling factors $\mathrm{s}$, $\mathrm{si}$, and $\mathrm{sr}$ for the next iteration. Line 46 determines the rescaled polynomial coefficient at order $n+1$. Lines 47-59 represent an iteration step \eqref{eq:dkmatparam} in rescaled form, and accumulation of terms for the $\cof{\overline{r}_{i,j}}_{i,j\in\cof{0,\ldots,N-1}}$ from \eqref{eq:rmatcoeffdef}. Finally, lines 60-69 implement the convergence test.}\label{algo:ch_and_chk_coeff}
\end{algorithm}
\end{figure*}

The pseudo code in Alg.~\ref{algo:ch_and_chk_coeff} illustrates how the recurrence relations \eqref{eq:dkmatparam} can be used to incorporate the computation of the coefficients $\cof{\bar{r}_{i,j}}_{i,j\in\cof{0,\ldots,N-1}}$ for \eqref{eq:rmatcoeffdef} in the algorithm for computing the $\cof{\bar{r}_{i}}_{i\in\cof{0,\ldots,N-1}}$ for \eqref{eq:chfresult}, shown in Alg.~\ref{algo:ch_coeff}. 

In Alg.~\ref{algo:ch_and_chk_coeff} the use of the rescaling procedure discussed in Sec.~\ref{sssec:stabiteration} to improve numerical stability is deeper incorporated than in Alg.~\ref{algo:ch_coeff}, since the stopping criterion in Alg.~\ref{algo:ch_and_chk_coeff} is no-longer based on convergence of the coefficients $\cof{\bar{r}_0,\ldots,\bar{r}_{N-1}}$ themselves, but more simply on the convergence of the running sum
\[
C\of{k}=\sum\limits_{n=0}^{k} r_{n}\,\tilde{\alpha}_n\ ,
\]
with the $\tilde{\alpha}_n$ from \eqref{eq:vectorchiterregb}. The algorithm stops after $n_{\mathrm{max}}$ iteration if within numerical precision $C\of{n_{\mathrm{max}}+1}=C\of{n_{\mathrm{max}}}$.

We conclude this section by establishing the connection between the coefficients $\cof{\bar{r}_{i,j}}_{i,j\in\cof{0,\ldots,N-1}}$ from \eqref{eq:rmatcoeffdef} and the $\cof{\dd \bar{r}_{i}}_{i\in\cof{0,\ldots,N-1}}$ from \eqref{eq:chfdiff}. To do this, we note that we can write the $\bar{r}_l$ from \eqref{eq:chfresult} in terms of the companion matrix $A$ from \eqref{eq:companionmat} as
\[
\bar{r}_l=\sum\limits_{n=0}^{\infty} r_n \of{A^n}\indices{^l_0}\ ,
\]
and consequently
\[
\dd\bar{r}_l=\sum\limits_{n=0}^{\infty} r_n \dd\of{A^n}\indices{^l_0}\ .\label{eq:drbaramat}
\]
We are interested in the components of the differential \eqref{eq:drbaramat} with respect to the basis given by the components of $\dd U$, i.e. 
\[
\partd{\bar{r}_l}{U\indices{^a_b}}\dd U\indices{^a_b}=\partd{\bar{r}_l}{A\indices{^c_d}}\underbrace{\partd{A\indices{^c_d}}{U\indices{^a_b}}}_{\mathclap{-\delta^{N-1}_d \partd{c_c}{U\indices{^a_b}}}} \dd U\indices{^a_b}\ .\label{eq:drbaramatcomp}
\]
For the derivatives of $\bar{r}_l$ with respect to the components of $A$ one finds:
\begin{multline}
\partd{\bar{r}_l}{A\indices{^c_d}}=\\
\sum\limits_{i,j=0}^{N-1}\bof{\underbrace{\sum\limits_{m=0}^{\infty}\,r_{m+1}\,\sum\limits_{k=0}^{m}\,a_{m-k,i}\,a_{k,j}}_{\bar{r}_{i,j}}}\of{A^i}\indices{^d_0}\of{A^j}\indices{^l_c}\\
=\sum\limits_{i,j=0}^{N-1}\,\bar{r}_{i,j}\,\of{A^i}\indices{^d_0}\of{A^j}\indices{^l_c}\ ,\label{eq:drbarda}
\end{multline}
where $\cof{\bar{r}_{i,j}}_{i,j\in\cof{0,\ldots,N-1}}$ are the same coefficients as in \eqref{eq:rmatcoeffdef}. The latter follows from the fact that the characteristic polynomial coefficients for the matrix $A$ are the same as those for the matrix $U$; hence the $\cof{a_{n,i}}_{n\in\mathbb{N}_0,i\in\cof{0,\ldots,N-1}}$ and the $\cof{\bar{r}_{i,j}}_{i,j\in\cof{0,\ldots,N-1}}$ in \eqref{eq:drbarda} and \eqref{eq:rmatcoeffdef} are computed from the same characteristic polynomial coefficients, and therefore respectively the same.
Plugging now \eqref{eq:drbarda} into \eqref{eq:drbaramatcomp} yields
\begin{multline}
\partd{\bar{r}_l}{U\indices{^a_b}}=-\sum\limits_{k=0}^{N-1}\partd{\bar{r}_l}{A\indices{^k_{N-1}}}\partd{c_k}{U\indices{^a_b}}\\
=-\sum\limits_{k=0}^{N-1}\bof{\sum\limits_{i,j=0}^{N-1}\bar{r}_{i,j}\underbrace{\of{A^i}\indices{^{N-1}_0}}_{\delta^i_{N-1}}\underbrace{\of{A^j}\indices{^l_k}}_{a_{j+k,l}}}\,\partd{c_k}{U\indices{^a_b}}\\
=-\sum\limits_{k=0}^{N-1}\bof{\underbrace{\sum\limits_{j=0}^{N-1}\bar{r}_{N-1,j}\,a_{j+k,l}}_{\tilde{R}_{l,k}}}\,\partd{c_k}{U\indices{^a_b}}\\
=-\sum\limits_{k=0}^{N-1}\tilde{R}_{l,k}\,\partd{c_k}{U\indices{^a_b}}\ ,\label{eq:drbarkmatdecomp}
\end{multline}
and by using \eqref{eq:charpolycoeffdiff} one finally arrives at:
\begin{multline}
\partd{\bar{r}_l}{U\indices{^a_b}}=\sum\limits_{k=0}^{N-1}\tilde{R}_{l,k}\sum\limits_{i=1}^{N-k}\frac{c_{k+i}}{i}\,\partd{p_i}{U\indices{^a_b}}\\
=\sum\limits_{i=0}^{N-1}\frac{1}{i+1}\sum\limits_{k=0}^{N-1-i} c_{k+i+1}\,\tilde{R}_{l,k}\,\underbrace{\partd{p_{i+1}}{U\indices{^a_b}}}_{\mathclap{\of{i+1}\,\of{P_i}\indices{^b_a}}}\\
=\sum\limits_{i=0}^{N-1}\bof{\underbrace{\sum\limits_{k=0}^{N-1-i} c_{k+i+1}\,\tilde{R}_{l,k}}_{R_{l,i}}}\,\of{P_i}\indices{^b_a}\\
=\sum\limits_{i=0}^{N-1}\,R_{l,i}\,\of{P_i}\indices{^b_a}\ .
\end{multline}

\section{Applications}\label{sec:applications}
An obvious application of the iterative Cayley-Hamilton method, described in the preceding Secs.~\ref{sec:method}-\ref{sec:implementation}, is the evaluation of matrix exponentials and their differentials (setting $r_n=1/n!$ in \eqref{eq:scalfunc} resp. \eqref{eq:chfresult}) in lattice gauge theory simulations, where matrix exponetiation is e.g. required to evolve the gauge fields in hybrid Monte Carlo (HMC)~\cite{Duane:1987de} updates or during gradient flow~\cite{Luscher:2010iy}. In combination with stout~\cite{Morningstar:2003gk} or HEX~\cite{Capitani:2006ni} smearing, these operatins involve also the matrix exponential differentials.

In this section, we will therefore first discuss the application of the iterative Cayley-Hamilton method to the computation of matrix exponentials and their differentials, in which case the method can be efficiently combined with "scaling and squaring"~\cite{Cleve:2006mfk} to improve convergence. After this, we will also discuss two less-obvious applications of the iterative Cayley-Hamilton method, namely the numerical determination of matrix logarithms of $\SU{N}$ matrices and the numerical computation of $\SU{N}$ one-link integrals. 

\subsection{Computation of matrix exponentials and their differentials}\label{ssec:matrixexp}
The power series representation of the exponential function has infinite radius of convergence, implying that it should in principle be straightforward to evaluate matrix exponentials with the method described in the previous section. When working with finite precision floating point arithmetic, one can, however, nevertheless run into difficulties. This can already happen in the scalar case, e.g. when computing the exponential of large negative numbers with the power-series approach: the terms initially grow rapidly in magnitude while their alternating signs lead to large cancellations and corresponding round-off errors. 

To avoid this kind of issue when evaluating the matrix exponential of some $U\in\mathbb{C}^{N\times N}$, one can make us of the fact that the exponential function satisfies,
\[
\exp\of{U}=\of{\exp\of{U/n}}^{n}\ .
\]
This leads to the so-called "scaling and squaring" method to improving the convergence of power-series-based matrix exponential compuations~\cite{Cleve:2006mfk}. The method works as follows: (a) determine the smallest $n=2^k$, $k\in\mathbb{N}_0$ for which $\abs{U}/n<c \leq 1$ with respect to some matrix norm $\abs{\cdot}$, (b) compute 
\[
V^{\of{0}}=\exp\of{U/n}\ ,\label{eq:reducedmatexp}
\]
and (c) determine $\exp\of{U}=V^{\of{k}}$ by iterative squaring:
\[
V^{\of{i}}=\sof{V^{\of{i-1}}}^2\ ,\quad i=1,\ldots,k\ .\label{eq:squaringiter}
\]
Since $\abs{U}/n<c$ with $0<c\leq 1$ the computation of the exponential in \eqref{eq:reducedmatexp} as power series does not suffer from rapidly growing terms and converges quickly. 

In terms of the $N\times N$ matrices $V^{\of{i}}$, each of the $k$ iterations in \eqref{eq:squaringiter} requires $\order{N^3}$ multiplication and addition operations. If \eqref{eq:reducedmatexp} is evaluated using the iterative Cayley-Hamilton method from the previous section, one gets
\[
V^{\of{0}}=\sum\limits{i=0}^{N-1}\,\bar{r}^{\of{0}}_{i}\ssof{\tilde{U}}\,\tilde{U}^{i}
\]
with the scaled matrix $\tilde{U}=U/n$. Using the product rule for Cayley-Hamilton decomposition coefficients, discussed in Sec.~\ref{sssec:productrule}, we can write the iterative squaring from \eqref{eq:squaringiter} directly in terms of these decomposition coefficients with respect to $\tilde{U}$ as:
\[
\bar{r}^{\of{n}}_{m}\ssof{\tilde{U}}=\sum\limits_{i,j=0}^{N-1}\,\bar{r}^{\of{n-1}}_{i}\ssof{\tilde{U}}\,\sof{\tilde{A}^i}\indices{^m_j}\,\bar{r}^{\of{n-1}}_{j}\ssof{\tilde{U}}\ ,\label{eq:chcoeffsquaringiter}
\]
where $\tilde{A}$ is the companion matrix for the characteristic polynomial of the scaled matrix $\tilde{U}$. Since the companion matrix is sparse with only $2$ non-zero entries per row, each squaring step in \eqref{eq:chcoeffsquaringiter} for $i=1,2\ldots,k$ can be carried out with only $\order{N^2}$ multiplication and addition operations (cf. Alg.~\ref{algo:ch_sq_coeffs} without gray code lines). Only the reconstruction of the final matrix
\[
\exp\of{U}=V^{\of{k}}=\sum\limits_{i=0}^{N-1}\,\bar{r}^{\of{k}}_{i}\ssof{\tilde{U}}\,\tilde{U}^{i}\ ,
\] 
from the coefficients $\cof{\smash{\bar{r}^{\of{k}}_{i}}}_{i\in\cof{0,\ldots,N-1}}$ and the pre-computed powers of $\tilde{U}$ requires $\order{N^3}$ operations.  

The scaling and squaring can also be performed for the computation of the derivatives of $\exp\of{U}$ with respect to the matrix components of $U$. Using \eqref{eq:rmatcoeffdef} we have 
\begin{multline}
\partd{\ssof{V^{\of{0}}}\indices{^a_b}}{\tilde{U}\indices{^c_d}}=\partd{\exp\ssof{\tilde{U}}\indices{^a_b}}{\tilde{U}\indices{^c_d}}\\
=\sum\limits_{i,j=0}^{N-1}\,\bar{r}^{\of{0}}_{i,j}\ssof{\tilde{U}}\,\ssof{\tilde{U}^i}\indices{^a_c}\,\ssof{\tilde{U}^j}\indices{^d_b}\ .
\end{multline}

Starting from
\begin{multline}
\partd{\exp\ssof{2^{n}\,\tilde{U}}\indices{^a_b}}{\ssof{2^{n}\,\tilde{U}}\indices{^c_d}}=\\
\frac{1}{2}\sum\limits_{e=0}^{N-1}\bof{\partd{\exp\ssof{2^{n-1}\tilde{U}}\indices{^a_e}}{\ssof{2^{n-1}\tilde{U}}\indices{^c_d}}\exp\ssof{2^{n-1}\tilde{U}}\indices{^e_b}\\
+\exp\ssof{2^{n-1}\tilde{U}}\indices{^a_e}\partd{\exp\ssof{2^{n-1}\tilde{U}}\indices{^e_b}}{\ssof{2^{n-1}\tilde{U}}\indices{^c_d}}}\ ,\label{eq:naiveksquaringiter}
\end{multline}
and making once more use of the product rule for Cayley-Hamilton coefficients from \ref{sssec:productrule}, one finds an iteration rule for computing the Cayley-Hamilton decomposition coefficients $\bar{r}^{\of{n}}_{i,j}\ssof{\tilde{U}}$, $i,j\in\cof{0,\ldots,N-1}$ for
\[
\partd{\exp\ssof{2^{n}\,\tilde{U}}\indices{^a_b}}{\ssof{2^{n}\,\tilde{U}}\indices{^c_d}}=\sum\limits_{i,j=0}^{N-1}\,\bar{r}^{\of{n}}_{i,j}\ssof{\tilde{U}}\,\ssof{\tilde{U}^i}\indices{^a_c}\,\ssof{\tilde{U}^j}\indices{^d_b}\ ,
\]
for $n=1,2,\ldots,k$, namely:
\begin{multline}
\bar{r}^{\of{n}}_{i,j}\ssof{\tilde{U}}=\frac{1}{2}\sum\limits_{l,m=0}^{N-1}\bof{\bar{r}^{\of{n-1}}_{i,l}\ssof{\tilde{U}}\sof{\tilde{A}^l}\indices{^j_m}\\
+\bar{r}^{\of{n-1}}_{j,l}\ssof{\tilde{U}}\sof{\tilde{A}^l}\indices{^i_m}}\,\bar{r}^{\of{n-1}}_{m}\ssof{\tilde{U}}\ .\label{eq:chkcoeffsquaringiter}
\end{multline}
Due to the sparseness of the companion matrix $\tilde{A}$, each iteration \eqref{eq:chkcoeffsquaringiter} requires only $\order{N^3}$ operations (cf. Alg.~\ref{algo:ch_sq_coeffs}).
%, while with \eqref{eq:naiveksquaringiter}, each iteration would require $\order{N^5}$ operations.

A C++ implementation of this method for computing matrix exponentials and their differentials using the iterative Cayley-Hamilton method in combination with scaling and squaring is included in~\cite{tobias_rindlisbacher_2024_10979006} and also available in the C++ lattice field theory programming framework for HPC simulations, "HILA"~\cite{HILA}.
\begin{algorithm}[!h]
\SetArgSty{textrm}
\SetFuncArgSty{textrm}
\SetStartEndCondition{ (}{)}{)}%
%\SetAlgoBlockMarkers{\{}{\}}%
%\SetKwFor{For}{for}{}{}%
%\SetKwFor{ForEach}{foreach}{}{}%
%\SetKwIF{If}{ElseIf}{Else}{if}{}{else if}{else}{}%
%\SetKwFor{While}{while}{}{}%
%\SetKwRepeat{Repeat}{repeat}{until}%
%\AlgoDisplayBlockMarkers\SetAlgoNoLine%
\SetKwFor{For}{for}{ \{}{}%
\SetKwFor{ForEach}{foreach}{}{}%
\SetKwIF{If}{ElseIf}{Else}{if}{ \{}{else if}{else \{}{}%
\SetKwFor{While}{while}{ \{}{}%
\SetKwRepeat{Repeat}{repeat \{}{until}%
\AlgoDisplayBlockMarkers%
\AlgoDisplayGroupMarkers%
\SetAlgoBlockMarkers{}{\}}%
\SetAlgoNoLine\SetAlgoNoEnd%
\SetKwArray{Pal}{a}
\SetKwArray{Kal}{ta}
\SetKwArray{Al}{rtot}
\SetKwArray{Crpl}{c}
\SetKwArray{Kh}{am}
\SetKwArray{Kmats}{rmtot}
\SetKwData{Cho}{tao}
\SetKwData{Ii}{i}
\SetKwData{Ij}{j}
\SetKwData{Ik}{k}
\SetKwData{Im}{m}
\SetKwData{Nmat}{N}
\SetKw{And}{$\&\&$}
\SetKw{Or}{$\|\|$}
\SetKw{Eq}{$==$}
\SetKw{SetTo}{$=$}
\SetKw{Neq}{$\neq$}
\SetKw{Gt}{$>$}
\SetKw{Lt}{$<$}
\SetKw{Geq}{$\geq$}
\SetKw{Leq}{$\leq$}
\SetKw{Mult}{$*$}
\SetKw{Div}{$/$}
\SetKw{Plus}{$+$}
\SetKw{Minus}{$-$}
\SetKw{PlusEq}{$+\!=$}
\SetKw{MinusEq}{$-\!=$}
\SetKw{MultEq}{$*\!=$}
\SetKw{DivEq}{$/\!=$}
\SetKw{Incr}{\!$++$}
\SetKw{Decr}{\!$--$}
\SetKw{Break}{break}
\SetKw{Itype}{itype}
\SetKw{Ctype}{ctype}
\SetKw{Ftype}{ftype}
\SetKwFunction{R}{r}
\SetKwFunction{Abs}{abs}
\SetKwFunction{Abssq}{abssq}
\SetKwFunction{Sqrt}{sqrt}
\SetKwInOut{Input}{input}
\SetKwInOut{Output}{output}
\SetKwInOut{Local}{local}
\Input{matrix size \Nmat\;\\
array \Crpl$=\cof{c_{0},\ldots,c_{N-1}}$\;\\
array \Al$=\cof{\bar{r}^{\of{n-1}}_{i}}_{i\in\cof{0,\ldots,N-1}}$\;\\
matrix \Kmats$=\cof{\bar{r}^{\of{n-1}}_{i,j}}_{i<j\in\cof{0,\ldots,N-1}}$\;
}
\Output{array \Al$=\cof{\bar{r}^{\of{n}}_{i}}_{i\in\cof{0,\ldots,N-1}}$\;\\
matrix \Kmats$=\cof{\bar{r}^{\of{n}}_{i,j}}_{i<j\in\cof{0,\ldots,N-1}}$\;}
\Local{temporary matrix of size \Nmat$\times$\Nmat: \Kh\;\\
temporary array of size \Nmat: \Kal, \Pal\;\\
temporary \Itype vars: \Ii,\ \Ij,\ \Ik\;\\
temporary \Ctype vars: \Cho\;
}
\BlankLine
\Cho \SetTo \Al{0}\; 
\For{\Ii\SetTo 0; \Ii\Lt\Nmat; \Ii\Incr}{
  \Kal{\Ii}\ \SetTo \Pal{\Ii}\ \SetTo \Al{\Ii}\;
  {\color{gray}\Kh{\Ii,\Ii}\ \SetTo 0.5 \Mult \Kmats{\Ii,\Ii}\;
  \Kmats{\Ii,\Ii}\ \SetTo 2.0 \Mult \Kh{0,\Ii}\ \Mult \Al{\Ii}\;
  \For{\Ij=\Ii \Plus 1; \Ij\Lt\Nmat; \Ij\Incr}{
    \Kh{\Ij,\Ii}\ \SetTo \Kh{\Ii,\Ij}\ \SetTo 0.5 \Mult \Kmats{\Ii,\Ij}\;
    \Kmats{\Ii,\Ij}\ \SetTo \Kh{0,\Ii}\ \Mult \Al{\Ij} + \Kh{0,\Ij}\ \Mult \Al{\Ii}\;
  }} 
  \Al{\Ii}\ \MultEq \Cho\;
}
\For{\Ik\SetTo 0; \Ik\Lt\Nmat; \Ik\Incr}{
  \Cho \SetTo \Al{\Nmat\Minus 1}\; 
  \For{\Ii\SetTo \Nmat\Minus 1; \Ii\Gt 0; \Ii\Decr}{
    \Pal{\Ii}\ \SetTo \Pal{\Ii\Minus 1}\ \Minus \Cho\Mult\Crpl{\Ii}\;
    \Al{\Ii}\ \PlusEq \Kal{\Ik}\ \Mult \Pal{\Ii}\;
    {\color{gray}\For{\Ij\SetTo \Ii; \Ij\Lt\Nmat; \Ij\Incr}{
       \Kmats{\Ii,\Ij}\,\PlusEq\!\Kh{\Ii,\Ik}\,\Mult\!\Pal{\Ij}\,\Plus\!\Kh{\Ik,\Ij}\,\Mult\!\Pal{\Ii}\;    
    }}
  }
  \Pal{0}\ \SetTo \Minus\!\Cho\Mult\Crpl{0}\;
  \Al{0}\ \PlusEq \Kal{\Ik}\ \Mult \Pal{0}\;
  {\color{gray}\For{\Ij\SetTo 0; \Ij\Lt\Nmat; \Ij\Incr}{
     \Kmats{0,\Ij}\,\PlusEq\!\Kh{0,\Ik}\,\Mult\!\Pal{\Ij}\,\Plus\!\Kh{\Ik,\Ij}\,\Mult\!\Pal{0}\;    
  }}
}
\caption{Implementation of the iteration steps \eqref{eq:chcoeffsquaringiter} and \eqref{eq:chkcoeffsquaringiter} to compute the Cayley-Hamilton decomposition coefficients $\overline{r}^{\of{n}}_{i}$, $\overline{r}^{\of{n}}_{i,j}$ $\forall\,i,j\in\cof{0,\ldots,N-1}$ of the $n$-times squared exponential and corresponding differential from the coefficients $\overline{r}^{\of{n-1}}_{i}$, $\overline{r}^{\of{n-1}}_{i,j}$ $\forall\,i,j\in\cof{0,\ldots,N-1}$ of the $\of{n-1}$-times squared exponential and corresponding differential. If the coefficients of the differential are not needed, the grey code lines can be skipped. Note that the array $\mathrm{\mathbf{rtot}}$ and the matrix $\mathrm{\mathbf{rmtot}}$ are used as both, input and output.}\label{algo:ch_sq_coeffs}
\end{algorithm}

\subsection{Computation of logarithms of $\SU{N}$ matrices}\label{ssec:logcomp}
In this section we discuss a method to determine the logarithm of $\SU{N}$ matrices, i.e.
\[
\log:\quad \SU{N}\to\su{N}\quad,\quad U\mapsto\omega=\log\of{U}\ ,
\]
with the imaginary parts of the eigenvalues of $\omega=\log\of{U}$ laying in the interval $\of{-\pi,\pi}$.

The method discussed in Sec.~\ref{sec:method} could in principle (in its shifted form \eqref{eq:chfresultshifted}) be used to compute the logarithm of a matrix $U$ from the Mercator series representation of the logarithm,
\[
\tilde{f}\of{x}=\log\of{1+x}=-\sum_{n=1}^{\infty}\,\frac{\of{-x}^n}{n}\ ,\label{eq:logpowerseries}
\]
setting $\tilde{U}=U-\id$ and $\log\of{U}=\tilde{f}\ssof{\tilde{U}}$.
However, as is well known, \eqref{eq:logpowerseries} holds only if $\abs{x}<1$, implying that the corresponding Cayley-Hamilton expansion for a $U\in\SU{N}$ would work only for a $U$ that is sufficiently close to $\id$ so that all eigenvalues of $\tilde{U}$ have modulus smaller than one.

An alternative approach to compute $\omega=\log\of{U}$ for $U\in\SU{N}$ can be reached by noting that the Lie algebra projection $A$ of $U$, given by
\[
A=P_{\su{N}}\of{U}=P_{\mathrm{ah}}\of{U}-\frac{\id}{N}\trace\of{P_{\mathrm{ah}}\of{U}}\ , \label{eq:liealgebraproj}
\]
with
\[
P_{\mathrm{ah}}\of{U}=\frac{1}{2}\of{U-U^{\dagger}}\ 
\]
being the anti-hermitian projection of $U$, can serve as an estimate for $\omega=\log\of{U}$, which gets the better the closer $U$ gets to $\id$.

One can then determine $\omega=\log\of{U}$ iteratively. Starting from $B_0=U$, $A_0=0$, one performs for subsequent $k=1,2,\ldots$ the operations 
\begin{subequations}\label{eq:logiteration}
\begin{align}
A_{k}&=A_{k-1}+P_{\su{N}}\of{B_{k-1}}\ ,\\
B_{k}&=U\,\exp\of{-A_{k}}\ ,
\end{align}
\end{subequations}
till for some $k=k_{\mathrm{max}}$, one has $\abs{P_{\su{N}}\of{B_{k-1}}}_1<\epsilon\,\abs{A_k}_1$, with $\abs{\cdot}_1$ being the element-wise matrix 1-norm and $\epsilon>0$ specifying the desired precision. The matrix exponential $\exp\of{-A_{k}}$ can be computed with the method described in Sec.~\ref{sec:implementation}, using the power series coefficients $r_n=1/n!$ for the exponential function. The number $k_{\mathrm{max}}$ of iterations over \eqref{eq:logiteration} needed to obtain $\omega=\log\of{U}=A_{k_{\mathrm{max}}}+\order{\epsilon}$ is typically $\sim N$ or lower (depending on how far $U$ is from $\id$) when working with double precision floating point arithmetcis (DPFPA), and $\epsilon$ set to $\epsilon=10\,N^2\,\epsilon_{\text{double}}$. Here $\epsilon_{\text{double}}$ is the "machine epsilon", i.e. the difference between 1.0 and the next larger representable number in DPFPA, and the factor of $10\,N^2$ is accounting for round-off errors in the components of $P_{\su{N}}\of{B_{k-1}}$, which accumulate when evaluating the 1-norm. A C++ implementations of this method for determining the matrix logarithm of $\SU{N}$ matrices is included in~\cite{tobias_rindlisbacher_2024_10979006} and also available in the "HILA" parallel programming framework~\cite{HILA}.

\subsection{Computation of $\SU{N}$ one-link integrals using Cayley-Hamilton}\label{ssec:onelinkint}
The Cayley-Hamilton theorem can be used to evaluate $\SU{N}$ one-link integrals of the form
\[
Z\of{S,S^{\dagger}}=\int\limits_{\mathclap{\SU{N}}}\dd U\,\e^{\trace\of{U S+U^{\dagger} S^{\dagger}}}\ ,\label{eq:onelinkint}
\]
where $S\in\mathbb{C}^{N\times N}$ and $\dd U$ is the $\SU{N}$ Haar measure. Our starting point is the formula by Brower et al.~\cite{Brower:1981vt},
\begin{multline}
Z\of{S,S^{\dagger}}=C\of{N} \sum_{l=-\infty}^{\infty}\e^{\ii\,l\,\theta} Z_{l}\of{S^{\dagger} S}\\
=C\of{N} \of{Z_0\of{S^{\dagger} S}+2\,\sum_{l=1}^{\infty}\cos\of{l\,\theta}\,Z_{l}\of{S^{\dagger} S}}\ ,\label{eq:onelinkint2}
\end{multline}
with
\[
C\of{N}=\prod\limits_{k=1}^{N-1}\,k!\label{eq:normconst}
\]
and
\[
Z_l\of{S^{\dagger} S}=\frac{\det\sof{z_i^{j} I_{j+l}\of{2\,z_i}}_{i,j\in\cof{0,\ldots,N-1}}}{\det\sof{x_{i}^{j}}_{i,j\in\cof{0,\ldots,N-1}}}\ ,\label{eq:ldetratio}
\]
where $\cof{x_i}_{i\in\cof{0,\ldots,N-1}}$ is the set of $N$ eigenvalues of the Hermitian matrix $M=S^{\dagger} S$ and we defined $z_i=\sqrt{x_i}$ $\forall\,i\in\cof{0,\ldots,N-1}$ and $\theta=\arg\of{\det\of{S}}$. The notation $\of{A_{i j}}_{i,j\in\cof{0,\ldots,N-1}}$ refers to a matrix with elements $A_{i j}$ and $\det\of{A_{i j}}_{i,j\in\cof{0,\ldots,N-1}}$ to the corresponding matrix determinant. The function $I_{\nu}\of{x}$ is the modified Bessel function of the first kind of order $\nu$, and 
\begin{multline}
V\sof{\cof{x}}=\sof{x_{i}^{j}}_{i,j\in\cof{0,\ldots,N-1}}\\
=\psmatrix{1 & x_0^1 & \cdots & x_0^{N-1}\\
1 & x_1^1 & \cdots & x_1^{N-1}\\
\vdots & \vdots &  & \vdots\vphantom{x_0^{N-1}}\\
1 & x_{N-1}^1 & \cdots & x_{N-1}^{N-1}}\label{eq:vandermondemat}
\end{multline}
is the Vandermonde matrix with respect to the support points $\cof{x}=\cof{x_i}_{i\in\cof{0,\ldots,N-1}}$.

While Eq.~\eqref{eq:onelinkint2} in combination with \eqref{eq:ldetratio} is more appropriate for the numerical evaluation of the one link integral \eqref{eq:onelinkint} than the derivative formula given in~\cite{Creutz:1978ub}, or the expansions in terms of Weingarten or Schur functions presented in~\cite{Borisenko:2018csw}, the computation of the $Z_{l}\of{S^{\dagger} S}$ via Eq.~\eqref{eq:onelinkint2} is only possible if $S^{\dagger} S$ has full rank and no coinciding eigenvalues, as otherwise the Vandermonde determinant in the denominator of Eq.~\eqref{eq:onelinkint2} would become zero. The latter case can be dealt with by modifying Eq.~\eqref{eq:ldetratio} to be
\[
\tilde{Z}_l\of{S^{\dagger} S}=\frac{\det\sof{\totd{\of{z_i^{j} I_{j+l}\of{2\,z_i}}}{z_i}{m_i}}_{i,j\in\cof{0,\ldots,N-1}}}{\det\sof{\totd{z_{i}^{2\,j}}{z_i}{m_i}}_{i,j\in\cof{0,\ldots,N-1}}}\ ,\label{eq:ldetratiomod}
\]
where the integer $m_i$ states how many times the value of $z_i$ appears among the $z_j$ with $j<i$. If all $z_i$, $i\in\cof{0,\ldots,N-1}$ are distinct one has $m_i=0$ $\forall i\in\cof{0,\ldots,N-1}$ and Eq.~\eqref{eq:ldetratiomod} redues to Eq.~\eqref{eq:ldetratio}. On the other hand, if e.g. $z_0=z_1$, one has $m_i=0$ for $i\neq 1$ and $m_1=1$, if $Z_0=z_1=z_2$, one has $m_i=0$ for $i\neq 1,2$ and $m_1=1$, $m_2=2$, and if e.g. $Z_0=z_1$ and $z_2=z_3$, one has $m_i=0$ for $i\neq 1,3$, and $m_1=1$, $m_3=1$, and so on.

In the remainder of this section, we will show how the Cayley-Hamilton theorem can be used to write Eq.~\eqref{eq:ldetratio} in terms of $S$ resp. $M$, without reference to the set $\cof{x_i}_{i\in\cof{0,\ldots,N-1}}$ of eigenvalues of $M$ or the corresponding square roots, $\cof{z_i}_{i\in\cof{0,\ldots,N-1}}$. Furthermore, the method we are about to describe for computing the $Z_{l}\of{S}$ from Eq.~\eqref{eq:ldetratio}, can be applied regardless of whether $M$ is degenerate or has repeated eigenvalues.  

To apply the Cayley-Hamilton method to the computation of Eq.~\eqref{eq:ldetratio}, we recall the power series representation of the modified Bessel functions of the first kind,
\[
I_{\nu}\of{2\,z}=\sum_{m=0}^{\infty}\frac{z^{\nu+2\,m}}{\of{\nu+m}!\,m!}\ .\label{eq:besselfuncpowerseries}
\]
From Eq.~\eqref{eq:besselfuncpowerseries} it follows that if we define
\[
B_{l,j}\of{x}=\sum_{m=0}^{\infty}\frac{l!\,x^{m+j}}{\of{l+m+j}!\,m!}=\sum_{n=j}^{\infty}\,r_{l,j,n}\,x^n\ ,
\]
with
\[
r_{l,j,n}=\frac{l!}{\of{l+n}!\,\of{n-j}!}\ ,\label{eq:modbesscoeffs}
\]
then 
\[
z^{\nu}I_{l+\nu}\of{2\,z}=\frac{z^{l}}{l!}B_{l,\nu}\of{z^2}\ ,
\]
and we can write Eq.~\eqref{eq:ldetratio} in the form
\[
Z_{l}\of{S^{\dagger} S}=\frac{\abs{\det\of{S}}^l}{l!^N}\,\frac{\det\sof{B_{l,j}\of{x_i}}_{i,j\in\cof{0,\ldots,N-1}}}{\det\sof{x_{i}^{j}}_{i,j\in\cof{0,\ldots,N-1}}}\ .
\]

Since $B_{l,j}\of{x}$ has infinite radius of convergence, we can replace the scalar argument $x$ by the matrix $M=S^{\dagger} S$ and apply the Cayley-Hamilton theorem to write
\[
B_{l,j}\of{M}=\sum_{k=0}^{N-1}\,\bar{r}_{l,j,k}\of{M}\,M^k\ .\label{eq:chexpofB}
\] 
In terms of the eigenvalues $\cof{x_i}_{i\in\cof{0,\ldots,N-1}}$ of $M=S^{\dagger} S$, Eq.~\eqref{eq:chexpofB} can also be written as:
\[
\psmatrix{B_{l,j}\of{x_0}\vphantom{x_0^{N-1}}\\
B_{l,j}\of{x_1}\vphantom{x_0^{N-1}}\\
\vdots\vphantom{x_0^{N-1}}\\
B_{l,j}\of{x_{N-1}}\vphantom{x_0^{N-1}}}=
\underbrace{\psmatrix{1 & x_0^1 & \cdots & x_0^{N-1}\\
1 & x_1^1 & \cdots & x_1^{N-1}\\
\vdots & \vdots &  & \vdots\vphantom{x_0^{N-1}}\\
1 & x_{N-1}^1 & \cdots & x_{N-1}^{N-1}}}_{V\ssof{\cof{x}}}
\psmatrix{\bar{r}_{l,j,0}\of{M}\vphantom{x_0^{N-1}}\\
\bar{r}_{l,j,1}\of{M}\vphantom{x_0^{N-1}}\\
\vdots\vphantom{x_0^{N-1}}\\
\bar{r}_{l,j,N-1}\of{M}\vphantom{x_0^{N-1}}}\ ,
\]
from which it follows that
\begin{multline}
\frac{\det\sof{B_{l,j}\of{x_i}}_{i,j\in\cof{0,\ldots,N-1}}}{\det\sof{x_{i}^{j}}_{i,j\in\cof{0,\ldots,N-1}}}\\
=\det\sof{V^{-1}\sof{\cof{x}}\sof{B_{l,j}\of{x_i}}_{i,j\in\cof{0,\ldots,N-1}}}\\
=\det\sof{\bar{r}_{l,j,i}\of{M}}_{i,j\in\cof{0,\ldots,N-1}}\ ,
\end{multline}
and therefore:
\begin{multline}
Z_{l}\of{S^{\dagger} S}=\frac{\abs{\det\of{S}}^l}{l!^N} \det\sof{\bar{r}_{l,j,i}\of{M}}_{i,j\in\cof{0,\ldots,N-1}}\\
=\frac{\abs{\det\of{S}}^l}{l!^N} \det\sof{R_{l}\of{M}}\ ,\label{eq:chzlofs}
\end{multline}
where on the last line we set
\[
R_{l}\of{M}=\sof{\bar{r}_{l,j,i}\of{M}}_{i,j\in\cof{0,\ldots,N-1}}\ .\label{eq:rlmdef}
\] 

The components $\bar{r}_{l,j,i}\of{M}$ of $R_{l}\of{M}$ are obtained as
\[
\bar{r}_{l,j,i}\of{M}=\sum_{n=j}^{\infty}\,r_{l,j,n}\,a_{n,i}\of{M}\ ,\label{eq:barrlji}
\]
with the $r_{l,j,n}$ defined in \eqref{eq:modbesscoeffs},
and the coefficients $\cof{a_{n,i}\of{M}}_{n\in\mathbb{N},i\in\cof{0,\ldots,N-1}}$ being determined with the method described in Sec.~\ref{sec:implementation}. Since it is desirable to compute all $\bar{r}_{l,j,i}\of{M}$ simultaneously so that each coefficient $a_{n,i}\of{M}$ is computed only once, the stopping criterion for the summation over $n$ in Alg.~\ref{algo:ch_coeff} is modified so that the summation stops as soon as all relevant $\bar{r}_{l,j,i}$ remain unchanged within machine precision for at least $\mathrm{nhl\_max}$ consecutive iterations. The number of relevant $l$-values, $l\in\cof{0,1,\ldots,l_{\mathrm{max}}-1}$ for which the coefficients \eqref{eq:barrlji} need to be computed is determined in advance by finding the smallest integer $l_{\mathrm{max}}>0$ for which within machine precision:
\[
\sum_{l=0}^{l_{\mathrm{max}}+1}\frac{\abs{\det\of{S}}^{l}}{l!^N}=\sum_{l=0}^{l_{\mathrm{max}}}\frac{\abs{\det\of{S}}^{l}}{l!^N} \ .
\]

By plugging Eq.~\eqref{eq:chzlofs} into Eq.~\eqref{eq:onelinkint2} we arrives at the following expression for $\SU{N}$ one-link integrals:
\begin{multline}
Z\of{S,S^{\dagger}}=C\of{N}\bof{\det\sof{R_{0}\ssof{S^{\dagger} S}}\,+\,\\
\sum_{l=1}^{\infty}\frac{\det^{l}\ssof{S}+\det^{l}\ssof{S^{\dagger}}}{l!^N} \det\sof{R_{l}\ssof{S^{\dagger} S}}}\ ,\label{eq:onelinkint3}
\end{multline}
from which it is not too difficult to determine the additional steps that would be required in the just described procedure to compute not only $Z\of{S,S^{\dagger}}$, but also its derivatives with respect to the components of $S$ or $S^{\dagger}$. The derivatives of $R_{l}\ssof{S^{\dagger} S}$ can be determined along the lines discussed in Sec.~\ref{ssec:computingderivterms}.

The determinant of the matrix $S$ and those of the matrices $R_l\of{S^{\dagger} S}$ are best computed via LU decomposition. It is also advisable to rescale the matrix $M=S^{\dagger} S$ for computing the $R_{l}\ssof{M}$ matrices, as discussed in Sec.~\ref{sssec:rescaledinputmat}, and to use the stabilization procedure for the Cayley-Hamilton iteration, described in Sec.~\ref{sssec:stabiteration}.

A C++ implementation of the here discussed method for computing $\SU{N}$ one-link integrals can be found under~\cite{tobias_rindlisbacher_2024_10979036}. The latter implementation is intended for applications in $\SU{N}$ lattice gauge theory in $d$ dimensions, so that $S$ is assumed to be a sum of $n_\text{stap.}=2\,\of{d-1}$ $\SU{N}$ staple matrices, multiplied by an inverse gauge coupling factor, $\beta/\of{2\,N}$. The rescaling factor $s$ for the matrix $M$ is then chosen to be $s=b^{-2}$ with 
\[
b=\of{d-1}\beta/N\ .\label{eq:effrfac}
\]
This choice of scaling factor results in a rescaled matrix $M$ whose largest eigenvalue has magnitude smaller than $1$, in which case one might want to modify the iteration stabilization procedure, discussed in Sec.~\ref{sssec:stabiteration}, so that it prevents the coefficients not just from growing too big, but also from becoming too small. 

The factor $b$ form \eqref{eq:effrfac} can furthermore be used to improve the condition numbers for the determinant computations of the matrices $R_{l}\of{M}$, $l\in\mathbb{N}_0$, namely by multiplying each element $\bar{r}_{l,j,i}\of{M}$ of \eqref{eq:rlmdef} by a corresponding factor $b^{i-j}$, using that:
\begin{multline}
\det\ssof{\bar{r}_{l,j,i}\of{M}}_{i,j\in\cof{0,\ldots,N-1}}\\
=\det\ssof{b^{i-j}\bar{r}_{l,j,i}\of{M}}_{i,j\in\cof{0,\ldots,N-1}}\ .
\end{multline}
Unfortunately, also after this balancing procedure the condition numbers of these matrices can still be large. 
 
\section{Numerical tests}\label{sec:numtests}

\subsection{Matrix exponentiation}\label{ssec:matrixexptest}
We compare two implementations of the iterative Cayley-Hamilton method for computing matrix exponentials with two other methods. One of the two Cayley-Hamilton implementations uses \textit{direct} input matrix re-scaling (CH+dSc), discussed in Sec.~\ref{sssec:rescaledinputmat}. This approach is equally applicable to the computation of other, more general matrix power series. The other Cayley-Hamilton implementation uses the \textit{scaling and squaring} approach (CH+Sc/Sq), discussed in in Sec.~\ref{ssec:matrixexp}. For better comparability, both methods use the same re-scaling factor, computed according to the scaling and squaring recipe, where for an input matrix $U$, the re-scaling factor $s$ is set to $s=2^{-k}$, with $k\in\mathbb{N}$ being the smallest integer for which $2^{-k}\,\abs{U}\leq 1$, with $\abs{\cdot}$ being the Frobenius norm. 

The methods to compare with are the adaptive Taylor series expansion with scaling and squaring (TS+Sc/Sq) and the 6th-order diagonal Padé approximation with scaling and squaring (Pd6+Sc/Sq) from~\cite{Cleve:2006mfk}, which will serve as reference. For the Taylor series method, the scaling and squaring re-scaling factor is computed in the same way as for the Cayley-Hamilton methods. For the Pd6+Sc/Sq method, it is determined with the maximum absolute column sum norm, $\abs{\cdot}_{\infty}$, instead of the Frobenius norm.

We use $\su{N}$ matrices to test and compare the different methods. For each $N\in\cof{2,3,\ldots,9,10,15,20}$ we prepare three sets, $S_r\of{N}$, of $10^4$ random $\su{N}$ matrices, sampled as Gaussian random vectors in $\mathbb{R}^{N^2-1}$ (spanned by the generators of $\su{N}$) and scaled to have fixed Frobenius norm $r$ with $r=\pi$, $3\,\pi$, and $r=4\,\pi$. The motivation for using $\pi$ as reference unit for the matrix norm is, that since the matrix norm is an upper bound for the magnitude of the largest eigenvalue of a matrix, it is for $r\leq\pi$ guaranteed that the matrices are located within the injectivity domain $I_{\exp}$ of the exponential map $\exp: \su{N} \to \SU{N}$ (cf.~\cite{rossmann2006lie},\cite[Sec.~II.B]{Rindlisbacher:2023qjg}), and as long as $s\,U\in I_{\exp}$, with $s\in\mathbb{R}$, $U\in\su{N}$, the matrix exponential $\exp\of{s\,U}\in \SU{N}$ does not show oscillatory behavior as function of $s$, which would give rise to cancellations and numerical loss of accuracy under \textit{direct} re-scaling in the CH+dSc method.

For the tests we pre-compute for every matrix $U\in S_{r}\of{N}$ its "exact" matrix exponential $\exp_{\mathrm{ex}}\of{U}$ to 50 digits precision, using the MatrixExp routine of Wolfram Mathematica (variable-order Padé approximation). The sets of matrix pairs $\tilde{S}_{r}\of{N}=\cof{\of{U,\exp_{\mathrm{ex}}\of{U}}\vert U\in S_{r}\of{N}}$ are then converted to double precision and stored as binary data for later use by the test programs (see~\cite{tobias_rindlisbacher_2024_10979006} for test data and programs). 

The test programs then measure for each method $m\in\cof{\text{CH+Sc/Sq},\text{CH+dSc},\text{TS+Sc/Sq},\text{Pd6+Sc/Sq}}$ and set of matrices $\tilde{S}_r\of{N}$ the time $t_m$ required to compute 100 times the matrix exponential of all $U\in S_r\of{N}$, as well as the relative errors
\[
\epsilon_m\of{U}=\frac{\abs{\exp_{m}\of{U}-\exp_{\mathrm{ex}}\of{U}}}{\abs{\exp_{\mathrm{ex}}\of{U}}}\ .\label{eq:relexperror}
\]

\begin{figure*}[htb]
\begin{minipage}[t]{0.3\linewidth}
\vspace{0pt}
\centering
\includegraphics[height=1.05\linewidth,keepaspectratio,right]{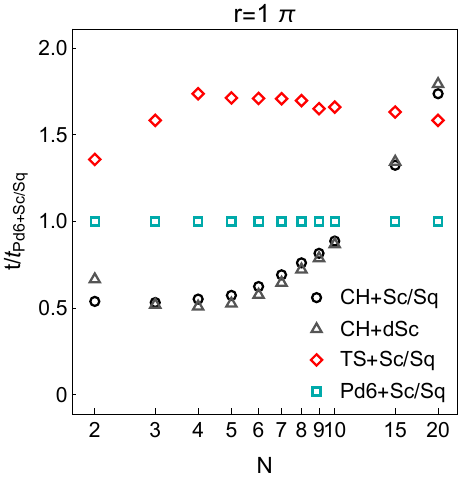}
\end{minipage}\hfill
\begin{minipage}[t]{0.3\linewidth}
\vspace{0pt}
\centering
\includegraphics[height=1.05\linewidth,keepaspectratio,right]{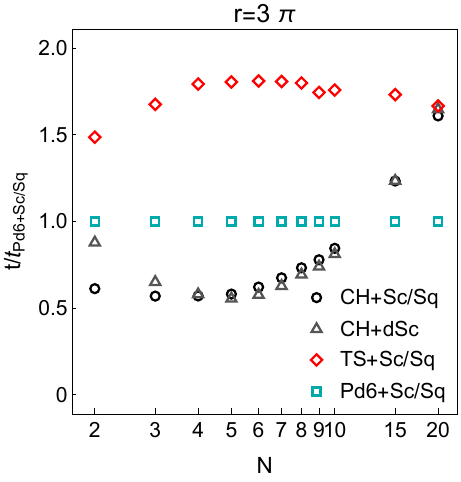}
\end{minipage}\hfill
\begin{minipage}[t]{0.3\linewidth}
\vspace{0pt}
\centering
\includegraphics[height=1.05\linewidth,keepaspectratio,right]{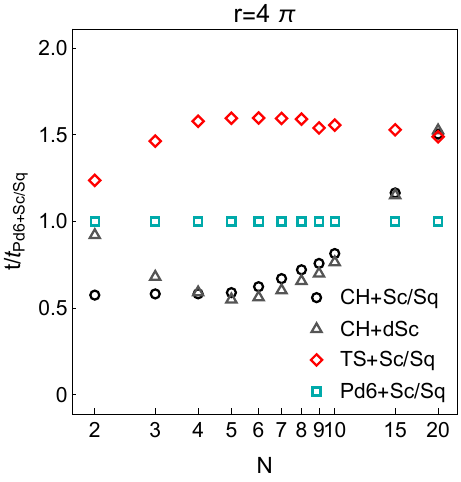}\end{minipage}\\[5pt]
\begin{minipage}[t]{0.3\linewidth}
\vspace{0pt}
\centering
\includegraphics[height=1.05\linewidth,keepaspectratio,right]{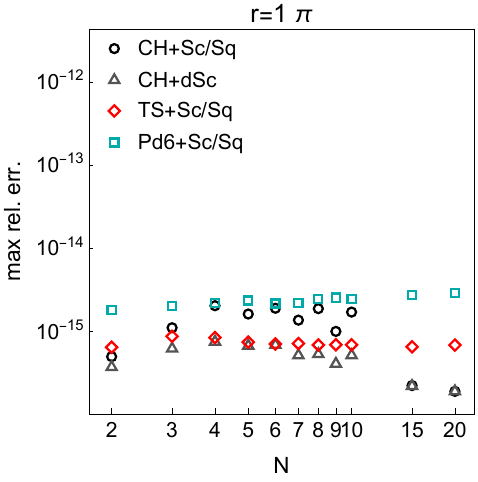}
\end{minipage}\hfill
\begin{minipage}[t]{0.3\linewidth}
\vspace{0pt}
\centering
\includegraphics[height=1.05\linewidth,keepaspectratio,right]{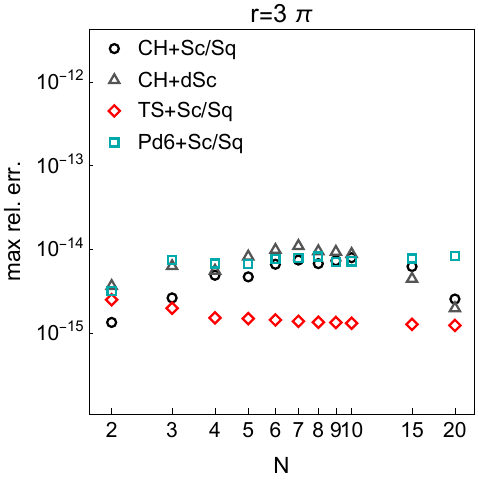}
\end{minipage}\hfill
\begin{minipage}[t]{0.3\linewidth}
\vspace{0pt}
\centering
\includegraphics[height=1.05\linewidth,keepaspectratio,right]{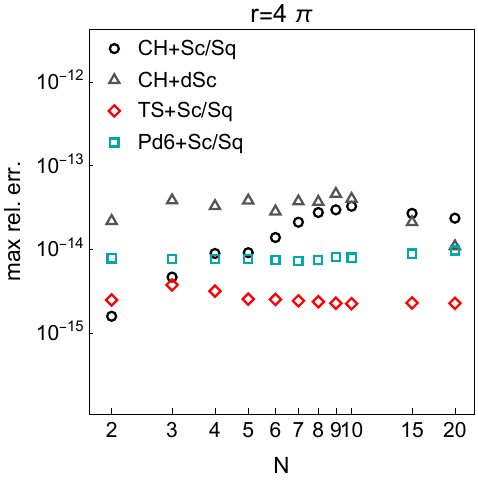}
\end{minipage}\\[5pt]
\caption{Comparison of computational cost (top) and numerical accuracy (bottom) of different methods to compute the matrix exponential of $\su{N}$ matrices for $N\in\cof{2,3,\ldots,9,10,15,20}$. The compared methods are: \textit{iterative Cayley-Hamilton with "scaling and squaring"} (CH+Sc/Sq) discussed in Sec.~\ref{ssec:matrixexp}, \textit{iterative Cayley-Hamilton with direct rescaling} (CH+dSc) discussed in Sec.~\ref{ssec:roundofferrors}, adaptive order \textit{Taylor series evaluation with "scaling and squaring"} (TS+Sc/Sq), and \textit{6th-order diagonal Padé approximation with "scaling and squaring"} (Pd6+Sc/Sq) from~\cite{Cleve:2006mfk}. The different columns show results for different sets of $10^4$ random $\su{N}$ matrices, sampled as Gaussian random vectors in $\mathbb{R}^{N^2-1}$ (spanned by the generators of $\su{N}$) and scaled so that their Frobenius norms have value $r=\pi$ (left), $r=3\,\pi$ (center), and $r=4\,\pi$ (right). The computational cost is defined as the time required to repeat with a given method 100 times the computation of the matrix exponentials of all $10^4$ matrices of a given set. The timing results are normalized by the Pd6+Sc/Sq timing result. The relative error is defined as the Frobenius norm of the difference between the result for the exponential of a given matrix computed with the method under consideration and the corresponding exact result, divided by the Frobenius norm of the exact result.}
\label{fig:numtests}
\end{figure*}

For stable performance, the tests have been carried out on a node of the HPC cluster "Mahti" of \textit{CSC - IT Center for Science, Finland}, in exclusive mode (no resources shared with other jobs). The results are summarized in Fig.~\ref{fig:numtests}: the first row shows the timing results for the different methods as ratios $t_m/t_{\text{Pd6+Sc/Sq}}$, while the second row shows the largest relative error observed with each method while computing the matrix exponentials of all $10^4$ matrices in each set, i.e., $\max_{U\in S_r\of{N}}\of{\epsilon_m\of{U}}$. 

As can be seen, up to $N\approx 10$ both iterative Cayley-Hamilton methods outperform the Pd6+Sc/Sq method while providing comparable accuracies. With increasing Frobenius norm $r$ of the input matrices, the accuracy of the CH+dSc method decreases faster than the accuracies of the other methods due to the aforementioned oscillatory behavior of the matrix exponential under re-scaling of the input matrix when not within the injectivity domain $I_{\exp}$ of the exponential map $\exp: \su{N} \to \SU{N}$. Also the relative error for the other methods grows with increasing $r$ due to the increasing number of squaring steps required after the exponential of the re-scaled matrix has been computed to machine precision. For the CH+Sc/Sq method the largest relative error grows faster with $r$ than for the TS+Sc/Sq and Pd6+Sc/Sq methods if $N>5$, but more slowly than for the CH+dSc method. 

Since the accuracies of the two iterative Cayley-Hamilton methods appear to drop below the accuracy of the Pd6+Sc/Sq method only for $r>3\pi$, while being significantly faster than the latter if $N<10$, the Cayley-Hamilton methods are well suited for applications in lattice gauge theory simulations. The matrix exponentials that need to be computed in such simulations are typically exponentials of $\su{N}$ matrices with $N\leq 10$ and which are typically of moderate Frobenius norm due to small pre-factors like small smearing coefficients or small time steps required by the used HMC or gradient flow integration schemes.

\subsection{HMC simulation with stout smearing}\label{ssec:hmcstouttest}

\begin{figure*}[htb]
\begin{minipage}[t]{0.325\linewidth}
\vspace{0pt}
\centering
\begin{minipage}[t]{0.96\linewidth}
\vspace{0pt}
\centering
\includegraphics[height=1.01\linewidth,keepaspectratio,right]{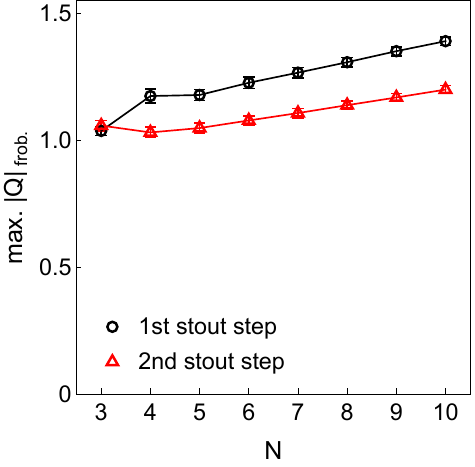}
\end{minipage}
\caption{Maximum Frobenius norms observed for the $Q_{\mu}^{\of{n}}\of{x}$ matrices from \eqref{eq:stoutqmatrix}, of which the matrix exponentials and their derivatives need to be computed during the HMC force computation for a two-step stout smeared (isotropc smearing parameter $c=0.15$) Wilson gauge action for $\SU{N}$. The simulations have been carried out on $V=16^4$ lattices with $\beta$ set for each $N$ to the corresponding $N_t=6$ pseudo critical values of the unsmeared Wilson gauge action.}
\label{fig:numstouttestmatsize}
\end{minipage}\hfill
\begin{minipage}[t]{0.65\linewidth}
\vspace{0pt}
\centering
\begin{minipage}[t]{0.48\linewidth}
\vspace{0pt}
\centering
\includegraphics[height=1.01\linewidth,keepaspectratio,right]{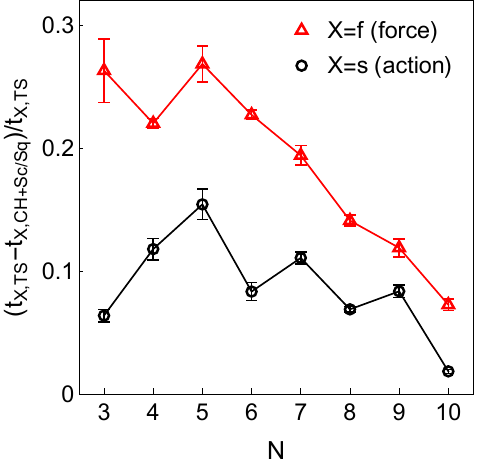}
\end{minipage}\hfill
\begin{minipage}[t]{0.48\linewidth}
\vspace{0pt}
\centering
\includegraphics[height=1.01\linewidth,keepaspectratio,right]{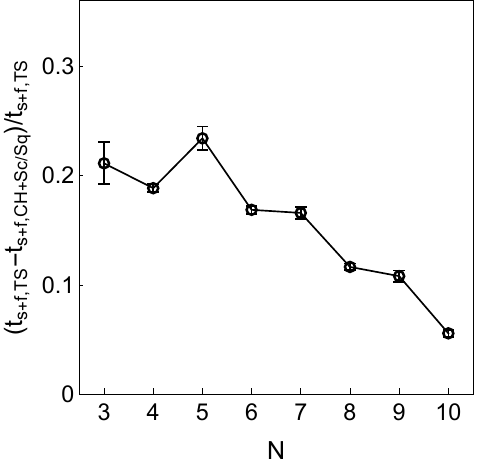}
\end{minipage}
\caption{Comparison of computational cost of gauge force computation in HMC simulations of pure $\SU{N}$ gauge theories with two-step stout smeared (isotropic smearing parameter $c=0.15$) Wilson gauge actions, using the iterative Cayley-Hamilton method with scaling and squaring (CH+Sc/Sq) and the adaptive Taylor series method (TS) for computing the involved matrix exponentials and their derivatives. In the left-hand panel $t_{\mathrm{X,TS}}$ and $t_{\mathrm{X,CH+Sc/Sq}}$ refer to the computer times required to compute the smeared action (X=s) and force (X=f), using the TS resp. CH+Sc/Sq method to evaluate the exponentials in \eqref{eq:stoutsmearedgaugelink} and the derivatives of the exponentials in \eqref{eq:stoutsmearedgaugeforcecomp}. The right-hand panel shows a comparison of the total computational cost (s+f) with the two methods. The observed net speedup with the (CH+Sc/Sq)-method is 19-24\% for $N\in\cof{3,4,5}$. Simulations have been carried out on $V=16^4$ lattices with $\beta$ set for each $N$ to the corresponding $N_t=6$ pseudo critical values of the unsmeared Wilson gauge action.}
\label{fig:numstouttests}
\end{minipage}
\end{figure*}

To test the iterative Cayley-Hamilton method for computing matrix exponentials in a use case where also the differentials of the exponentials are required, we set up HMC simulations of pure $\SU{N}$ lattice gauge theories, governed by a $n_s$-step stout smeared Wilson gauge action $S_G\ssfof{V^{\of{n_s}}\ssfof{U}}$. Here $U=\cof{U_{\mu}\of{x}}_{x,\mu}$ is the fundamental lattice gauge field with the link variables $U_{\mu}\of{x}\in\SU{N}$ $\forall x,\mu$ being defined on oriented links $\of{x,\mu}$, with $x$ being the coordinate of a lattice site and $\mu=1,\ldots,d$ the direction in which the link leaves the site. 
The Wilson gauge $S_G$ in terms of a gauge field $U$ is given by, 
\[
S_{G}\sfof{U}=\frac{\beta}{N}\sum\limits_{x,\mu<\nu}\repart\trace\sof{\id-U_{\mu\nu}\of{x}}\ ,\label{eq:smearedgaugeaction}
\]
where, $\beta=2\,N/g_0^2$, is the inverse bare gauge coupling and $U_{\mu\nu}\of{x}$ are the plaquette variables,
\[
U_{\mu\nu}\of{x}=U_{\mu}\of{x}\,U_{\nu}\of{x+\hat{\mu}}\,U_{\mu}^{\dagger}\of{x+\hat{\nu}} \,U_{\nu}^{\dagger}\of{x}\ .\label{eq:plaqvar}
\]
The stout smearing procedure used to map the fundamental gauge field $U$ to the smeared gauge field $V^{\of{n_s}}\ssfof{U}=\cof{V_{\mu}^{\of{n_s}}\of{x}}_{x,\mu}$, consisting of the smeared link variables $V_{\mu}^{\of{n_s}}\of{x}\in\SU{N}$ $\forall x,\mu$, is described in Appendix~\ref{app:stoutsmearing}. 

The simulation program is implemented using the "HILA" parallel programming framework and available under~\cite[applications/suN\_hmc\_stout\_comp]{HILA}. The computation of the HMC force of the smeared action is done with the CH+Sc/Sq method. It makes use of Algs.~\ref{algo:ch_and_chk_coeff} and \ref{algo:ch_sq_coeffs} to pre-compute the Cayley-Hamilton decomposition coefficients for the matrix exponential differentials while carrying out the matrix exponentiatials for the forward smearing from \eqref{eq:stoutsmearedgaugelink}. These coefficients are then used in the force computation in \eqref{eq:stoutsmearedgaugeforcecomp} to compute the terms that involve the derivatives of the exponential without requiring renewed Cayley-Hamilton iterations. For comparison, the stout smearing procedure is also implemented using the adaptive Taylor series method (TS). 

With both methods, we perform simulations for $N=3,\ldots,10$ on $V=16^4$ lattices, using $n_s=2$ stout smearing steps and isotropic smearing parameter $c=0.15$. The inverse gauge coupling $\beta$ is for each $N$ set to the $N_t=6$ pseudo critical value of the unsmeared Wilson gauge action. Since $\beta$ multiplies in our simulations a smeared action, the latter choice is to some extent arbitrary, but has the effect that the unsmeared plaquette action has an almost constant value across all considered values of $N$, i.e., $\avof{\repart\trace\of{\id-U_{\mu\nu}}/N}\approx 0.8$ for all $N=3,\ldots,10$).

Fig.~\ref{fig:numstouttestmatsize} shows for each $N$ the maximum of the in the simulations observed Frobenius norms $r=\ssabs{Q_{\mu}^{\of{n}}\of{x}}$ of the matrices \eqref{eq:stoutqmatrix} of which the exponentials are required in \eqref{eq:stoutsmearedgaugelink}. As can be seen, the maximum values range from $r=1.0$ to $r=1.4$. All matrices are therefore far from reaching values $r=3\pi$ where the accuracy of the CH+Sc/Sq method would start to be lower than that of the Pd6+Sc/Sq method, used as reference in the previous section (cf. Fig.~\ref{fig:numtests}). 

In Fig.~\ref{fig:numstouttests} the computational costs for the gauge force computation with the CH+Sc/Sq and the TS method are compared. The left-hand panel shows the cost reductions achieved with the CH+Sc/Sq method (in comparison with the TS method) in the computation of the smeared action (black circles), which involves the forward smearing operation from \eqref{eq:stoutsmearedgaugelink}, and in the computation of the smeared force (red triangles) according to \eqref{eq:stoutsmearedgaugeforcecomp}. 
As mentioned above, with the CH+Sc/Sq method, the Cayley-Hamilton decomposition coefficents for the differentials of the exponentials, required to evaluate \eqref{eq:stoutsmearedgaugeforcecomp}, are computed along with the exponentials themselves in \eqref{eq:stoutsmearedgaugelink}. This increases slightly the computational cost of the action computation while decreasing the cost for the force computation with the CH+Sc/Sq method. The right-hand panel of Fig.~\ref{fig:numstouttests} therefore also shows the reduction in computational cost for the whole smeared HMC force computation, which amounts to 19-24\% for $N\in\cof{3,4,5}$. Note that the timing results include computational steps (e.g. staple and staple force computation) which are independent of the used method for evaluating matrix exponentials and their differentials. The speedup is therefore smaller than what one might have expected from the results in Fig.~\ref{fig:numtests}.  Error bars represent standard deviations of timing results from 10 independent single node runs (2 $\times$ AMD Rome 7H12 CPU with 128 cores in total) on the "Mahti" cluster of \textit{CSC - IT Center for Science, Finland}.

\section{Summary}\label{sec:summary}
The Cayley-Hamilton theorem has been used in Sec.~\ref{sec:method} to derive an iterative process that allows for the efficient computation of matrix power series $f\of{U}$ on arbitrary square matrices $U$. For moderately sized matrices the method is significantly faster than a naive evaluation of $f\of{U}$ and, if necessary, also allows for the simultaneous computation of the derivatives of $f\of{U}$ with respect to the components of $U$. 

A possible implementation of the basic algorithm was discussed in Sec.~\ref{sec:implementation}. 

The algorithm is well suited for use in HMC simulations~\cite{Duane:1987de} of lattice gauge theories. For example, to efficiently perform the matrix exponentials required for stout~\cite{Morningstar:2003gk} and HEX~\cite{Capitani:2006ni} smearing, and the computation of the HMC force for smeared actions. When applied to the computation of matrix exponentials and their derivatives, the method can efficiently be combined with scaling and squaring~\cite{Cleve:2006mfk}.

In Sec.~\ref{sec:applications} we proposed two further applications of the iterative Cayley-Hamilton method, namely to ease the numerical computation of $\SU{N}$ one link integrals and to compute matrix logarithms of $\SU{N}$ matrices. 

In Sec.~\ref{sec:numtests} we applied two versions of the iterative Cayley-Hammilton method for computing matrix exponentials (with and without scaling and squaring) to the problem of mapping $\su{N}$ matrices to $\SU{N}$, and compared their performances to those of two other matrix exponentiation methods: a 6th-order diagonal Padé method with scaling and squaring~\cite{Cleve:2006mfk} and an adaptive Taylor series method with scaling and squaring. As expected the Taylor series method was computationally the most expensive one, requiring about 1.4-1.8 times the computer time of the Padé method. However, the Taylor series method turned out to produce slightly more accurate results. The iterative Cayley-Hamilton methods clearly outperformed the Padé method for $N\leq 10$. For $N<7$ the scaling and squaring version of the Cayley-Hamilton method required only about half the computer time of the Padé method while being equally accurate as long as the Frobenius norms of the input matrices did not significantly exceed the value $\sim 3\,\pi$ (cf. Fig.~\ref{fig:numtests}). For input matrices of larger norms, the accuracies of the Cayley-Hamilton methods started to drop faster than those of the other methods. For applications in lattice gauge theory simulations, the latter is of little concern, since the matrices whose exponentials need to be computed in such applications have typically relatively small norms due to small pre-factors like small smearing coefficients or small HMC or gradient flow integration time steps. We performed HMC simulations with two-step stout smeared Wilson gauge actions for $\SU{N}$, $N=2,3,\ldots,10$. With isotropic stout smearing coefficient $c=0.15$, the largest observed Frobenius norms of the matrices, of which the exponentials (and their differentials) had to be computed in these simulations, were between 1.0 ($N=3$) and 1.4 ($N=10$), far from the limit where the Cayley-Hamilton methods would start to loose accuracy.   

\section{Acknowledgements}\label{sec:acknowledgments}

The author acknowledges support from the Swiss National Science Foundation (SNSF) through the grant no.~210064. The author thanks CSC - IT Center for Science, Finland, for computational resources.

\appendix
\section{Stout smearing procedure}\label{app:stoutsmearing}

The $n_s$-step stout smeared~\cite{Morningstar:2003gk} link variables $V^{\of{n_s}}_{\mu}\of{x}$ are defined iteratively, through the relation
\[
V^{\of{n}}_{\mu}\of{x}=\exp\sof{Q^{\of{n-1}}_{\mu}\of{x}}\,V^{\of{n-1}}_{\mu}\of{x}\quad\forall\,x,\mu \label{eq:stoutsmearedgaugelink}
\]
for $n=1,\ldots,n_s$ with $V^{\of{0}}_{\mu}\of{x}=U_{\mu}\of{x}$ $\forall x,\mu$. The matrices which are in \eqref{eq:stoutsmearedgaugelink} argument to the exponential function are given by
\[
Q^{\of{n}}_{\mu}\of{x}=P_{\su{N}}\sof{C^{\of{n}}_{\mu}\of{x}\,V^{\of{n}}_{\mu}{}^{\dagger}\of{x}}\ ,\label{eq:stoutqmatrix}
\]
where $P_{\su{N}}$ is the Lie algebra projection introduced in \eqref{eq:liealgebraproj} and $C^{\of{n}}_{\mu}\of{x}$ is the weighted sum of smeared (inverse) staples around the link $\of{x,\mu}$:
\begin{multline}
C^{\of{n}}_{\mu}\of{x}=\\
\sum\limits_{\nu\neq\mu} \rho_{\mu\nu}\of{x}\bof{V^{\of{n}}_{\nu}\of{x}\,V^{\of{n}}_{\mu}\of{x+\hat{\nu}}\,V^{\of{n}}_{\nu}{}^{\dagger}\of{x+\hat{\mu}}\\
+V^{\of{n}}_{\nu}{}^{\dagger}\of{x-\hat{\nu}}\,V^{\of{n}}_{\mu}\of{x-\hat{\nu}}\,V^{\of{n}}_{\nu}\of{x-\hat{\nu}+\hat{\mu}}}\ .\label{eq:stoutsmearedstaplesum}
\end{multline}
We will set $\rho_{\mu\nu}\of{x}=c\ ,\forall\,x,\mu,\nu$, i.e. use isotropic and homogeneous smearing.

To carry out HMC updates with the smeared gauge action, we need to compute also the corresponding gauge force. We start by writing
\begin{multline}
\delta^{I}_{y,\rho} S_{G}\sfof{V^{\of{n_s}}\sfof{U}}=\\
-\frac{\beta}{N}\sum\limits_{x,\mu}\trace\sof{\Pi_{\mu}^{\of{n_s}}\of{x}\,\ssof{\delta^{I}_{y,\rho} V^{\of{n_s}}_{\mu}\of{x}}\,V^{\of{n_s}}_{\mu}{}^{\dagger}\of{x}}\label{eq:smearedactionvari}
\end{multline}
where $\delta^{I}_{y,\rho}$ denotes the derivative of the fundamental link $U_{\rho}\of{y}$ in the direction of the $I$-th $\su{N}$ generator, so that $\delta^{I}_{y,\rho}U_{\mu}\of{x}=\delta_{x,y}\delta_{\mu,\rho}\,T^I\,U_{\mu}\of{x}$, and $\Pi_{\mu}^{\of{n}}\of{x}\in\su{N}$ is the force acting on the smeared link $V^{\of{n}}_{\mu}\of{x}$. At smearing level $n=n_s$ the expression for $\Pi_{\mu}^{\of{n_s}}\of{x}$ is essentially the same as the expression for $Q^{\of{n_s}}_{\mu}\of{x}$ with a negative sign and the factor of $\rho_{\mu\nu}\of{x}$ dropped from the expression for $C^{\of{n_s}}_{\mu}\of{x}$ in \eqref{eq:stoutsmearedstaplesum}, i.e. 
\[
\Pi_{\mu}^{\of{n_s}}\of{x}=-\sum\limits_{\nu\neq\mu}\partd{Q_{\mu}^{\of{n_s}}\of{x}}{\rho_{\mu\nu}\of{x}}\ .
\]
To compute the force at subsequent lower smearing levels and ultimately the one acting on the fundamental link $U_{\mu}\of{x}$, we
substitute \eqref{eq:stoutsmearedgaugelink} into the variation of the smeared link $V^{\of{n}}_{\mu}\of{x}$ to obtain the relation
\begin{multline}
\delta^{I}_{y,\rho} V^{\of{n}}_{\mu}\of{x} = \exp\ssof{Q^{\of{n-1}}_{\mu}\of{x}}\,\delta^{I}_{y,\rho} V^{\of{n-1}}_{\mu}\of{x}\\
+\sof{\underbrace{\delta^{I}_{y,\rho}\exp\ssof{Q^{\of{n-1}}_{\mu}\of{x}}}_{\mathclap{\partd{\exp\ssof{Q^{\of{n-1}}_{\mu}\of{x}}}{\ssof{Q^{\of{n-1}}_{\mu}\of{x}}\indices{^a_b}}\sum\limits_{z,\nu}\partd{\ssof{Q^{\of{n-1}}_{\mu}\of{x}}\indices{^a_b}}{\ssof{V_{\nu}^{\of{n-1}}\of{z}}\indices{^c_d}}\,\ssof{\delta^{I}_{y,\rho}V_{\nu}^{\of{n-1}}\of{z}}\indices{^c_d}}}}\,V^{\of{n-1}}_{\mu}\of{x}\ .\label{eq:stoutlinkvari}
\end{multline}
If we now substitute \eqref{eq:stoutlinkvari} back into \eqref{eq:smearedactionvari}, we find that we can write
\begin{multline}
\delta^{I}_{y,\rho} S_{G}\sfof{V^{\of{n_s}}\sfof{U}}=\\
-\frac{\beta}{N}\sum\limits_{x,\mu}\trace\sof{\Pi_{\mu}^{\of{n_s}}\of{x}\,\ssof{\delta^{I}_{y,\rho} V^{\of{n_s}}_{\mu}\of{x}}\,V^{\of{n_s}}_{\mu}{}^{\dagger}\of{x}}\\
=-\frac{\beta}{N}\sum\limits_{x,\mu}\trace\sof{\Pi_{\mu}^{\of{n_s-1}}\of{x}\,\ssof{\delta^{I}_{y,\rho} V^{\of{n_s-1}}_{\mu}\of{x}}\,V^{\of{n_s-1}}_{\mu}{}^{\dagger}\of{x}}\\
\vdots\\
=-\frac{\beta}{N}\sum\limits_{x,\mu}\trace\sof{\Pi_{\mu}^{\of{0}}\of{x}\,\ssof{\delta^{I}_{y,\rho} U_{\mu}\of{x}}\,U_{\mu}^{\dagger}\of{x}}\\
=-\frac{\beta}{N}\trace\sof{\Pi_{\rho}^{\of{0}}\of{y}\,T^{I}}\ ,\label{eq:smearedactionvari2}
\end{multline}
with the force terms at subsequent lower smearing levels defined iteratively by
\begin{multline}
\ssof{\Pi^{\of{n-1}}_{\mu}\of{x}}\indices{^d_c}=\\
\sof{\exp\ssof{-Q^{\of{n-1}}_{\mu}\of{x}}\,\Pi^{\of{n}}_{\mu}\of{x}\,\exp\ssof{Q^{\of{n-1}}_{\mu}\of{x}}}\indices{^d_c}\\
+\sum\limits_{z,\nu}\bfof{\!\trace\bof{\!\!\exp\ssof{-Q^{\of{n-1}}_{\nu}\of{z}}\,\Pi^{\of{n}}_{\nu}\of{z}\,\partd{\exp\ssof{Q^{\of{n-1}}_{\nu}\of{z}}}{\ssof{Q^{\of{n-1}}_{\nu}\of{z}}\indices{^a_b}}\!}\\
\cdot\partd{\ssof{Q^{\of{n-1}}_{\nu}\of{z}}\indices{^a_b}}{\ssof{V_{\mu}^{\of{n-1}}\of{x}}\indices{^c_d}}}\ ,\label{eq:stoutsmearedgaugeforcecomp}
\end{multline}
for $n=n_s,\ldots,1$, and where \eqref{eq:stoutsmearedgaugelink} was used to write
\[
V^{\of{n-1}}_{\mu}\of{x}\,V^{\of{n}}_{\mu}{}^{\dagger}\of{x}=\exp\sof{-Q^{\of{n-1}}_{\mu}\of{x}}\ .
\]

\bibliography{biblio}

\begin{thebibliography}{19}%
\makeatletter
\providecommand \@ifxundefined [1]{%
 \@ifx{#1\undefined}
}%
\providecommand \@ifnum [1]{%
 \ifnum #1\expandafter \@firstoftwo
 \else \expandafter \@secondoftwo
 \fi
}%
\providecommand \@ifx [1]{%
 \ifx #1\expandafter \@firstoftwo
 \else \expandafter \@secondoftwo
 \fi
}%
\providecommand \natexlab [1]{#1}%
\providecommand \enquote  [1]{``#1''}%
\providecommand \bibnamefont  [1]{#1}%
\providecommand \bibfnamefont [1]{#1}%
\providecommand \citenamefont [1]{#1}%
\providecommand \href@noop [0]{\@secondoftwo}%
\providecommand \href [0]{\begingroup \@sanitize@url \@href}%
\providecommand \@href[1]{\@@startlink{#1}\@@href}%
\providecommand \@@href[1]{\endgroup#1\@@endlink}%
\providecommand \@sanitize@url [0]{\catcode `\\12\catcode `\$12\catcode
  `\&12\catcode `\#12\catcode `\^12\catcode `\_12\catcode `\%12\relax}%
\providecommand \@@startlink[1]{}%
\providecommand \@@endlink[0]{}%
\providecommand \url  [0]{\begingroup\@sanitize@url \@url }%
\providecommand \@url [1]{\endgroup\@href {#1}{\urlprefix }}%
\providecommand \urlprefix  [0]{URL }%
\providecommand \Eprint [0]{\href }%
\providecommand \doibase [0]{https://doi.org/}%
\providecommand \selectlanguage [0]{\@gobble}%
\providecommand \bibinfo  [0]{\@secondoftwo}%
\providecommand \bibfield  [0]{\@secondoftwo}%
\providecommand \translation [1]{[#1]}%
\providecommand \BibitemOpen [0]{}%
\providecommand \bibitemStop [0]{}%
\providecommand \bibitemNoStop [0]{.\EOS\space}%
\providecommand \EOS [0]{\spacefactor3000\relax}%
\providecommand \BibitemShut  [1]{\csname bibitem#1\endcsname}%
\let\auto@bib@innerbib\@empty
%</preamble>
\bibitem [{\citenamefont {Morningstar}\ and\ \citenamefont
  {Peardon}(2004)}]{Morningstar:2003gk}%
  \BibitemOpen
  \bibfield  {author} {\bibinfo {author} {\bibfnamefont {C.}~\bibnamefont
  {Morningstar}}\ and\ \bibinfo {author} {\bibfnamefont {M.~J.}\ \bibnamefont
  {Peardon}},\ }\bibfield  {title} {\bibinfo {title} {{Analytic smearing of
  SU(3) link variables in lattice QCD}},\ }\href
  {https://doi.org/10.1103/PhysRevD.69.054501} {\bibfield  {journal} {\bibinfo
  {journal} {Phys. Rev. D}\ }\textbf {\bibinfo {volume} {69}},\ \bibinfo
  {pages} {054501} (\bibinfo {year} {2004})},\ \Eprint
  {https://arxiv.org/abs/hep-lat/0311018} {arXiv:hep-lat/0311018} \BibitemShut
  {NoStop}%
\bibitem [{\citenamefont {Capitani}\ \emph {et~al.}(2006)\citenamefont
  {Capitani}, \citenamefont {Durr},\ and\ \citenamefont
  {Hoelbling}}]{Capitani:2006ni}%
  \BibitemOpen
  \bibfield  {author} {\bibinfo {author} {\bibfnamefont {S.}~\bibnamefont
  {Capitani}}, \bibinfo {author} {\bibfnamefont {S.}~\bibnamefont {Durr}},\
  and\ \bibinfo {author} {\bibfnamefont {C.}~\bibnamefont {Hoelbling}},\
  }\bibfield  {title} {\bibinfo {title} {{Rationale for UV-filtered clover
  fermions}},\ }\href {https://doi.org/10.1088/1126-6708/2006/11/028}
  {\bibfield  {journal} {\bibinfo  {journal} {JHEP}\ }\textbf {\bibinfo
  {volume} {11}},\ \bibinfo {pages} {028}},\ \Eprint
  {https://arxiv.org/abs/hep-lat/0607006} {arXiv:hep-lat/0607006} \BibitemShut
  {NoStop}%
\bibitem [{\citenamefont {Hasenfratz}\ and\ \citenamefont
  {Knechtli}(2001)}]{Hasenfratz:2001hp}%
  \BibitemOpen
  \bibfield  {author} {\bibinfo {author} {\bibfnamefont {A.}~\bibnamefont
  {Hasenfratz}}\ and\ \bibinfo {author} {\bibfnamefont {F.}~\bibnamefont
  {Knechtli}},\ }\bibfield  {title} {\bibinfo {title} {{Flavor symmetry and the
  static potential with hypercubic blocking}},\ }\href
  {https://doi.org/10.1103/PhysRevD.64.034504} {\bibfield  {journal} {\bibinfo
  {journal} {Phys. Rev. D}\ }\textbf {\bibinfo {volume} {64}},\ \bibinfo
  {pages} {034504} (\bibinfo {year} {2001})},\ \Eprint
  {https://arxiv.org/abs/hep-lat/0103029} {arXiv:hep-lat/0103029} \BibitemShut
  {NoStop}%
\bibitem [{\citenamefont {Hasenfratz}\ \emph {et~al.}(2007)\citenamefont
  {Hasenfratz}, \citenamefont {Hoffmann},\ and\ \citenamefont
  {Schaefer}}]{Hasenfratz:2007rf}%
  \BibitemOpen
  \bibfield  {author} {\bibinfo {author} {\bibfnamefont {A.}~\bibnamefont
  {Hasenfratz}}, \bibinfo {author} {\bibfnamefont {R.}~\bibnamefont
  {Hoffmann}},\ and\ \bibinfo {author} {\bibfnamefont {S.}~\bibnamefont
  {Schaefer}},\ }\bibfield  {title} {\bibinfo {title} {{Hypercubic smeared
  links for dynamical fermions}},\ }\href
  {https://doi.org/10.1088/1126-6708/2007/05/029} {\bibfield  {journal}
  {\bibinfo  {journal} {JHEP}\ }\textbf {\bibinfo {volume} {05}},\ \bibinfo
  {pages} {029}},\ \Eprint {https://arxiv.org/abs/hep-lat/0702028}
  {arXiv:hep-lat/0702028} \BibitemShut {NoStop}%
\bibitem [{\citenamefont {DeGrand}\ \emph {et~al.}(2012)\citenamefont
  {DeGrand}, \citenamefont {Shamir},\ and\ \citenamefont
  {Svetitsky}}]{DeGrand:2012qa}%
  \BibitemOpen
  \bibfield  {author} {\bibinfo {author} {\bibfnamefont {T.}~\bibnamefont
  {DeGrand}}, \bibinfo {author} {\bibfnamefont {Y.}~\bibnamefont {Shamir}},\
  and\ \bibinfo {author} {\bibfnamefont {B.}~\bibnamefont {Svetitsky}},\
  }\bibfield  {title} {\bibinfo {title} {{SU(4) lattice gauge theory with
  decuplet fermions: Schr\"odinger functional analysis}},\ }\href
  {https://doi.org/10.1103/PhysRevD.85.074506} {\bibfield  {journal} {\bibinfo
  {journal} {Phys. Rev. D}\ }\textbf {\bibinfo {volume} {85}},\ \bibinfo
  {pages} {074506} (\bibinfo {year} {2012})},\ \Eprint
  {https://arxiv.org/abs/1202.2675} {arXiv:1202.2675 [hep-lat]} \BibitemShut
  {NoStop}%
\bibitem [{\citenamefont {DeGrand}\ and\ \citenamefont
  {Liu}(2016)}]{DeGrand:2016pur}%
  \BibitemOpen
  \bibfield  {author} {\bibinfo {author} {\bibfnamefont {T.}~\bibnamefont
  {DeGrand}}\ and\ \bibinfo {author} {\bibfnamefont {Y.}~\bibnamefont {Liu}},\
  }\bibfield  {title} {\bibinfo {title} {{Lattice study of large $N_c$ QCD}},\
  }\href {https://doi.org/10.1103/PhysRevD.94.034506} {\bibfield  {journal}
  {\bibinfo  {journal} {Phys. Rev. D}\ }\textbf {\bibinfo {volume} {94}},\
  \bibinfo {pages} {034506} (\bibinfo {year} {2016})},\ \bibinfo {note}
  {[Erratum: Phys.Rev.D 95, 019902 (2017)]},\ \Eprint
  {https://arxiv.org/abs/1606.01277} {arXiv:1606.01277 [hep-lat]} \BibitemShut
  {NoStop}%
\bibitem [{\citenamefont {Hudspith}(2014)}]{Hudspith:2014rma}%
  \BibitemOpen
  \bibfield  {author} {\bibinfo {author} {\bibfnamefont {R.~J.}\ \bibnamefont
  {Hudspith}},\ }\bibfield  {title} {\bibinfo {title} {{Conjugate Directions in
  Lattice Landau and Coulomb Gauge Fixing}},\ }\href
  {https://doi.org/10.22323/1.214.0048} {\bibfield  {journal} {\bibinfo
  {journal} {PoS}\ }\textbf {\bibinfo {volume} {LATTICE2014}},\ \bibinfo
  {pages} {048} (\bibinfo {year} {2014})},\ \Eprint
  {https://arxiv.org/abs/1412.2807} {arXiv:1412.2807 [hep-lat]} \BibitemShut
  {NoStop}%
\bibitem [{\citenamefont {Francis}\ \emph {et~al.}(2020)\citenamefont
  {Francis}, \citenamefont {Fritzsch}, \citenamefont {L\"uscher},\ and\
  \citenamefont {Rago}}]{Francis:2019muy}%
  \BibitemOpen
  \bibfield  {author} {\bibinfo {author} {\bibfnamefont {A.}~\bibnamefont
  {Francis}}, \bibinfo {author} {\bibfnamefont {P.}~\bibnamefont {Fritzsch}},
  \bibinfo {author} {\bibfnamefont {M.}~\bibnamefont {L\"uscher}},\ and\
  \bibinfo {author} {\bibfnamefont {A.}~\bibnamefont {Rago}},\ }\bibfield
  {title} {\bibinfo {title} {{Master-field simulations of O($a$)-improved
  lattice QCD: Algorithms, stability and exactness}},\ }\href
  {https://doi.org/10.1016/j.cpc.2020.107355} {\bibfield  {journal} {\bibinfo
  {journal} {Comput. Phys. Commun.}\ }\textbf {\bibinfo {volume} {255}},\
  \bibinfo {pages} {107355} (\bibinfo {year} {2020})},\ \Eprint
  {https://arxiv.org/abs/1911.04533} {arXiv:1911.04533 [hep-lat]} \BibitemShut
  {NoStop}%
\bibitem [{\citenamefont {Cleve}\ and\ \citenamefont
  {Charles}(2006)}]{Cleve:2006mfk}%
  \BibitemOpen
  \bibfield  {author} {\bibinfo {author} {\bibfnamefont {M.}~\bibnamefont
  {Cleve}}\ and\ \bibinfo {author} {\bibfnamefont {V.~L.}\ \bibnamefont
  {Charles}},\ }\bibfield  {title} {\bibinfo {title} {{Nineteen Dubious Ways to
  Compute the Exponential of a Matrix, Twenty-Five Years Later}},\ }\href
  {https://doi.org/10.1137/S00361445024180} {\bibfield  {journal} {\bibinfo
  {journal} {SIAM Rev.}\ }\textbf {\bibinfo {volume} {45}},\ \bibinfo {pages}
  {3} (\bibinfo {year} {2006})}\BibitemShut {NoStop}%
\bibitem [{\citenamefont
  {Rindlisbacher}(2024{\natexlab{a}})}]{tobias_rindlisbacher_2024_10979006}%
  \BibitemOpen
  \bibfield  {author} {\bibinfo {author} {\bibfnamefont {T.}~\bibnamefont
  {Rindlisbacher}},\ }\href@noop {} {\bibinfo {title} {{C++ implementation of
  iterative Cayley-Hamilton method for computing matrix power series}}},\
  \bibinfo {howpublished} {doi:
  \href{https://doi.org/10.5281/zenodo.10979006}{10.5281/zenodo.10979006}}
  (\bibinfo {year} {2024}{\natexlab{a}})\BibitemShut {NoStop}%
\bibitem [{\citenamefont {Duane}\ \emph {et~al.}(1987)\citenamefont {Duane},
  \citenamefont {Kennedy}, \citenamefont {Pendleton},\ and\ \citenamefont
  {Roweth}}]{Duane:1987de}%
  \BibitemOpen
  \bibfield  {author} {\bibinfo {author} {\bibfnamefont {S.}~\bibnamefont
  {Duane}}, \bibinfo {author} {\bibfnamefont {A.~D.}\ \bibnamefont {Kennedy}},
  \bibinfo {author} {\bibfnamefont {B.~J.}\ \bibnamefont {Pendleton}},\ and\
  \bibinfo {author} {\bibfnamefont {D.}~\bibnamefont {Roweth}},\ }\bibfield
  {title} {\bibinfo {title} {{Hybrid Monte Carlo}},\ }\href
  {https://doi.org/10.1016/0370-2693(87)91197-X} {\bibfield  {journal}
  {\bibinfo  {journal} {Phys. Lett. B}\ }\textbf {\bibinfo {volume} {195}},\
  \bibinfo {pages} {216} (\bibinfo {year} {1987})}\BibitemShut {NoStop}%
\bibitem [{\citenamefont {L\"uscher}(2010)}]{Luscher:2010iy}%
  \BibitemOpen
  \bibfield  {author} {\bibinfo {author} {\bibfnamefont {M.}~\bibnamefont
  {L\"uscher}},\ }\bibfield  {title} {\bibinfo {title} {{Properties and uses of
  the Wilson flow in lattice QCD}},\ }\href
  {https://doi.org/10.1007/JHEP08(2010)071} {\bibfield  {journal} {\bibinfo
  {journal} {JHEP}\ }\textbf {\bibinfo {volume} {08}},\ \bibinfo {pages}
  {071}},\ \bibinfo {note} {[Erratum: JHEP 03, 092 (2014)]},\ \Eprint
  {https://arxiv.org/abs/1006.4518} {arXiv:1006.4518 [hep-lat]} \BibitemShut
  {NoStop}%
\bibitem [{HIL()}]{HILA}%
  \BibitemOpen
  \href@noop {} {\bibinfo {title} {Hila lattice simulation framework}},\
  \bibinfo {howpublished} {\url{https://github.com/CFT-HY/HILA}}\BibitemShut
  {NoStop}%
\bibitem [{\citenamefont {Brower}\ \emph {et~al.}(1981)\citenamefont {Brower},
  \citenamefont {Rossi},\ and\ \citenamefont {Tan}}]{Brower:1981vt}%
  \BibitemOpen
  \bibfield  {author} {\bibinfo {author} {\bibfnamefont {R.}~\bibnamefont
  {Brower}}, \bibinfo {author} {\bibfnamefont {P.}~\bibnamefont {Rossi}},\ and\
  \bibinfo {author} {\bibfnamefont {C.-I.}\ \bibnamefont {Tan}},\ }\bibfield
  {title} {\bibinfo {title} {{The External Field Problem for {QCD}}},\ }\href
  {https://doi.org/10.1016/0550-3213(81)90046-8} {\bibfield  {journal}
  {\bibinfo  {journal} {Nucl. Phys. B}\ }\textbf {\bibinfo {volume} {190}},\
  \bibinfo {pages} {699} (\bibinfo {year} {1981})}\BibitemShut {NoStop}%
\bibitem [{\citenamefont {Creutz}(1978)}]{Creutz:1978ub}%
  \BibitemOpen
  \bibfield  {author} {\bibinfo {author} {\bibfnamefont {M.}~\bibnamefont
  {Creutz}},\ }\bibfield  {title} {\bibinfo {title} {{ON INVARIANT INTEGRATION
  OVER SU(N)}},\ }\href {https://doi.org/10.1063/1.523581} {\bibfield
  {journal} {\bibinfo  {journal} {J. Math. Phys.}\ }\textbf {\bibinfo {volume}
  {19}},\ \bibinfo {pages} {2043} (\bibinfo {year} {1978})}\BibitemShut
  {NoStop}%
\bibitem [{\citenamefont {Borisenko}\ \emph {et~al.}(2020)\citenamefont
  {Borisenko}, \citenamefont {Voloshyn},\ and\ \citenamefont
  {Chelnokov}}]{Borisenko:2018csw}%
  \BibitemOpen
  \bibfield  {author} {\bibinfo {author} {\bibfnamefont {O.}~\bibnamefont
  {Borisenko}}, \bibinfo {author} {\bibfnamefont {S.}~\bibnamefont
  {Voloshyn}},\ and\ \bibinfo {author} {\bibfnamefont {V.}~\bibnamefont
  {Chelnokov}},\ }\bibfield  {title} {\bibinfo {title} {{Su$(N)$ Polynomial
  Integrals and Some Applications}},\ }\href
  {https://doi.org/10.1016/S0034-4877(20)30015-X} {\bibfield  {journal}
  {\bibinfo  {journal} {Rept. Math. Phys.}\ }\textbf {\bibinfo {volume} {85}},\
  \bibinfo {pages} {129} (\bibinfo {year} {2020})},\ \Eprint
  {https://arxiv.org/abs/1812.06069} {arXiv:1812.06069 [hep-lat]} \BibitemShut
  {NoStop}%
\bibitem [{\citenamefont
  {Rindlisbacher}(2024{\natexlab{b}})}]{tobias_rindlisbacher_2024_10979036}%
  \BibitemOpen
  \bibfield  {author} {\bibinfo {author} {\bibfnamefont {T.}~\bibnamefont
  {Rindlisbacher}},\ }\href@noop {} {\bibinfo {title} {{C++ implementation of a
  SU(N) one-link integrator using the iterative Cayley-Hamilton method}}},\
  \bibinfo {howpublished} {doi:
  \href{https://doi.org/10.5281/zenodo.10979036}{10.5281/zenodo.10979036}}
  (\bibinfo {year} {2024}{\natexlab{b}})\BibitemShut {NoStop}%
\bibitem [{\citenamefont {Rossmann}(2006)}]{rossmann2006lie}%
  \BibitemOpen
  \bibfield  {author} {\bibinfo {author} {\bibfnamefont {W.}~\bibnamefont
  {Rossmann}},\ }\href@noop {} {\emph {\bibinfo {title} {Lie Groups}}},\ Oxford
  Graduate Texts in Mathematics\ (\bibinfo  {publisher} {Oxford University
  Press},\ \bibinfo {address} {London, England},\ \bibinfo {year}
  {2006})\BibitemShut {NoStop}%
\bibitem [{\citenamefont {Rindlisbacher}\ \emph {et~al.}(2023)\citenamefont
  {Rindlisbacher}, \citenamefont {Rummukainen},\ and\ \citenamefont
  {Salami}}]{Rindlisbacher:2023qjg}%
  \BibitemOpen
  \bibfield  {author} {\bibinfo {author} {\bibfnamefont {T.}~\bibnamefont
  {Rindlisbacher}}, \bibinfo {author} {\bibfnamefont {K.}~\bibnamefont
  {Rummukainen}},\ and\ \bibinfo {author} {\bibfnamefont {A.}~\bibnamefont
  {Salami}},\ }\bibfield  {title} {\bibinfo {title}
  {{Bulk-transition-preventing actions for SU(N) gauge theories}},\ }\href
  {https://doi.org/10.1103/PhysRevD.108.114511} {\bibfield  {journal} {\bibinfo
   {journal} {Phys. Rev. D}\ }\textbf {\bibinfo {volume} {108}},\ \bibinfo
  {pages} {114511} (\bibinfo {year} {2023})},\ \Eprint
  {https://arxiv.org/abs/2306.14319} {arXiv:2306.14319 [hep-lat]} \BibitemShut
  {NoStop}%
\end{thebibliography}%

\end{document}